\documentclass[particles,article,accept,pdftex,moreauthors]{Definitions/mdpi}
\usepackage{amssymb,color,graphicx,bm,mathrsfs,color,amsmath,slashed,comment,aas_macros}

\newcommand{\be}{\begin{equation}}
\newcommand{\ee}{\end{equation}}
\newcommand{\bea}{\begin{eqnarray}}
\newcommand{\eea}{\end{eqnarray}}
\newcommand{\half}{\frac1 2}

\newcommand{\vectau}{{\bm \tau}}
\newcommand{\vecrho}{{\bm \rho}}
\newcommand{\ie}{{i.e.}}
\newcommand{\eg}{{e.g.}}
\newcommand{\MeV}{{\rm MeV}}
\newcommand{\fromto}{\leftrightarrow}

\input{acronym.input}

\usepackage[normalem]{ulem}  

\firstpage{361}
\pubvolume{5}
\issuenum{3}
\articlenumber{29}
\pubyear{2022}
\copyrightyear{2022}
\externaleditor{Academic Editor: {Nicolas Chamel} 
}
\datereceived{30 July 2022}
\dateaccepted{31 August 2022}
\datepublished{6 September 2022}
\hreflink{https://doi.org/10.3390/\linebreak particles5030029} 

\Title{Bulk Viscosity of Relativistic $npe\mu$ Matter in\linebreak Neutron-Star Mergers}

\TitleCitation{Bulk Viscosity of Relativistic $npe\mu$ Matter in Neutron-Star Mergers}

\Author{{Mark} 
Alford $^{1}$\href{https://orcid.org/0000-0001-9675-7005}{\orcidicon}, Arus Harutyunyan $^{2,3,}$*\href{https://orcid.org/0000-0002-3006-8618}{\orcidicon} and Armen Sedrakian $^{4,5}$\href{https://orcid.org/0000-0001-9626-2643}{\orcidicon}}

\AuthorNames{Mark Alford, Arus Harutyunyan and Armen Sedrakian}

\AuthorCitation{Alford, M.; Harutyunyan, A.; Sedrakian, A.}

\address{%
$^{1}$ \quad Department of {Physics,} 
Washington University, St.~Louis,
MO 63130, USA
\\
$^{2}$ \quad Byurakan Astrophysical Observatory, National Academy of
Sciences,  Byurakan 0213, Armenia
\\
$^{3}$ \quad Department of {Physics,} 
Yerevan State University,
Yerevan 0025, Armenia\\
$^{4}$ \quad Frankfurt Institute for Advanced Studies, D-60438
Frankfurt am Main, Germany
\\
$^{5}$ \quad Institute of Theoretical Physics, University of Wroc\l{}aw,
50-204 Wroc\l{}aw, Poland }

\corres{Correspondence: arus@bao.sci.am}

\abstract{We discuss the bulk viscosity of hot and dense $npe\mu$ matter
arising from weak-interaction direct Urca processes. We consider
two regimes of interest: (a) the neutrino-transparent regime with
$T\leq T_{\rm tr}$ ($T_{\rm tr}\simeq 5\div 10$~MeV is the
neutrino-trapping temperature); and (b) the neutrino-trapped
regime with $T\geq T_{\rm tr}$. Nuclear matter is modeled in
relativistic density functional approach with density-dependent
parametrization DDME2. The maximum of the bulk viscosity is
achieved at temperatures $T \simeq 5\div 6$~MeV in the
neutrino-transparent regime, then it drops rapidly at higher
temperatures where neutrino-trapping occurs. As an astrophysical
application, we estimate the damping timescales of density oscillations
by the bulk viscosity in neutron star mergers and find that, \eg,
at the oscillation frequency $f=10$~kHz, the damping will be very
efficient at temperatures $4\leq T\leq 7$~MeV where the bulk
viscosity might affect the evolution of the post-merger object.}

\keyword{bulk viscosity; weak processes; $npe\mu$ matter; binary neutron star mergers; damping of density oscillations }

\begin{document}

\section{Introduction}
\label{sec:intro}

The recent detections of gravitational waves from binary neutron-star
(BNS) mergers by the LIGO-Virgo collaboration motivate studies of the
transport properties of hot and dense nuclear matter. Numerical
simulations of BNS mergers predict intense emission of gravitational
waves during the first tens of milliseconds after the merger in the
kHz frequency range (see, \eg,
Refs.~\cite{Endrizzi2018,Most2019,Ciolfi2019,Tsokaros2019} for recent
simulations). The dissipation of matter flows in the
post-merger object might affect the gravitational wave spectra emitted during this
stage of BNS merger evolution. In particular, indications of bulk viscous dissipation were seen in a recent BNS simulation
incorporating beta equilibrating processes \cite{Most:2022yhe}, confirming earlier estimates of its
likely importance.

There exist extensive studies of the bulk viscosity of
neutron--proton--electron (hereafter $npe$) and $npe\mu$ (where $\mu$
stands for muons) matters in low-temperature (cold) neutron
stars~\cite{Sawyer1979ApJ,Sawyer1980ApJ,
Sawyer1989,Haensel1992PhRvD,Haensel2000,Haensel2001,Haensel2002,Dong2007,
Alford2010JPhG,Alford:2010jf,Kolomeitsev2015}. The bulk viscosity
of the dense matter at high temperatures (up to tens of MeV) was computed in
recent works which covered various regimes of temperature and density,
as well as neutrino trapping/un-trapping, in strongly interacting
hadronic matter~\cite{Alford2019a,Alford2019b,Alford2021a,Alford2021c}.

In this contribution, we review briefly the results of
Ref.~\cite{Alford2021c} for the bulk viscosity of the
neutrino-trapped, relativistic $npe\mu$ matter, as well as complement
them with novel results for the neutrino-transparent regime. The
impact of purely leptonic weak processes on the bulk viscosity is
discussed. We use the DDME2 parametrization~\cite{Lalazissis2005} of
relativistic density functional theory with density-dependent
couplings to model the background nuclear matter. {It provides
very reasonable values of such characteristics of symmetric nuclear
matter, as  the energy per nucleon $E_{\rm sat} = -16.14$~MeV and
compressibility $K_{\rm sat}~=~251.15$~MeV at saturation density
$n_{0}=0.152$ fm$^{-3}$, as well as characteristics of asymmetric
nuclear matter such as the symmetry energy
$E_{\rm sym} = 32.31$\,MeV and its slope $L_{\rm sym}= 51.27$
MeV. In our previous work~\cite{Alford2019b, Alford2021c} we used
the NL3 parameterization (along with the DDME2) where the couplings
are density-independent but meson fields acquire additional
self-interactions terms. This functional differs significantly from the DDME2
functional used in this study in the properties of
asymmetric matter with $E_{\rm sym} = 37.4$ and $L_{\rm sym}= 118.9$
MeV.  Specifically, these two functionals cover well the range for the
parameters $E_{\rm sym}$ and $L_{\rm sym}$ that have been inferred
from the PREX-II experiment by two alternative
{analysis~} 
\cite{Reed_2021, Reinhard_2021}. Thus, using our previous
results one can assess the impact of the variations of the important
characteristics of nuclear matter on the various quantities of
interest, such as bulk viscosity and damping time scales. A full
analysis of the sensitivity of the results on the input of various
density functionals goes beyond the present study.}

With the results obtained
for the bulk viscosity, we estimate the bulk viscous dissipation
timescales of density oscillations in BNS mergers.  For typical
oscillation frequencies $1\leq f\leq 10$~kHz the bulk viscous damping
timescales reach down to tens of milliseconds (at $n_B\simeq 3n_0$) or
milliseconds (at $n_B\simeq n_0$) at temperatures $4\leq T\leq 7$~MeV.
Here the bulk viscous damping can have a significant impact on the
initial phase of post-merger dynamics with a typical timescale
$\sim$10~ms. At high temperatures above the neutrino-trapping, the
bulk viscosity falls rapidly by orders of magnitude, and the damping
timescales become too long to affect the dynamics of BNS mergers.

This paper is organized as follows. In Section~\ref{sec:urca_rates}, we
discuss the weak processes in nuclear matter. In Section~\ref{sec:bulk},
we discuss the bulk viscosity produced by the Urca processes.
Section~\ref{sec:num_results} collects the numerical results for
the equilibration rates, the bulk viscosity, and the dissipation
damping timescales in the regimes of neutrino-transparent and
neutrino-trapped matter. Section~\ref{sec:conclusions} provides
a brief summary of our results. We work with natural (Gaussian)
units where $\hbar = c = k_B = 1$. 

\section{Weak Processes in Neutron Star Matter}
\label{sec:urca_rates}

We consider relativistic $npe\mu$ matter in the range of densities
$0.5n_0\leq n_B\leq 5 n_0$, where $n_0\simeq 0.152$ fm$^{-3}$ is the
nuclear saturation density and temperatures $1\leq T \simeq
100$~MeV. Neutrinos are trapped in the matter above the
neutrino-trapping temperature $T_{\rm tr}\simeq 5\div 10$~MeV and
un-trapped (free-streaming) below this temperature~\cite{Alford2018b}.

Consider now the simplest semi-baryonic $\beta$-equilibration
processes---the direct Urca processes of neutron decay and
lepton capture, respectively
\bea\label{eq:n_decay}
&& n\rightleftarrows p + l^-+\bar{\nu}_l,\\
\label{eq:l_capture}
&& p + l^-\rightleftarrows n+{\nu}_l,
\eea
where $l=\{e,\mu\}$ is electron or muon, $\nu_l$ is the
corresponding neutrino. In the $\nu$-transparent regime, these
processes proceed only in one direction from left to right as
neutrinos/anti-neutrinos can appear only in the final state.

If muons are present in matter, the following leptonic processes of
muon decay, neutrino~absorption and antineutrino~absorption,
respectively, may occur additionally
\bea\label{eq:reaction_L1}
&& \mu^- \rightleftarrows e^- +\bar{\nu}_e +{\nu}_\mu,\\
\label{eq:reaction_L2}
&& \mu^- +{\nu}_e \rightleftarrows e^-+ {\nu}_\mu,\\
\label{eq:reaction_L3}
&& \mu^- +\bar{\nu}_\mu  \rightleftarrows e^-+\bar{\nu}_e.
\eea

In the $\nu$-transparent regime, \eqref{eq:reaction_L2} and \eqref{eq:reaction_L3} are not possible, and \eqref{eq:reaction_L1} can only occur in the forward direction when the temperature is high enough to open up enough phase space around the muon and electron Fermi surfaces. We neglect modified-Urca-type processes involving electromagnetic interaction with a spectator particle \cite{Alford:2010jf}; these are suppressed by a factor of $\alpha^2$.

In $npe\mu$ matter the baryon number given by $n_B=n_n+n_p$ is
conserved. { The matter} is also charge neutral,
{\ie,} $n_p=n_e+n_\mu$. In the neutrino-trapped case the lepton numbers
$n_{L_l}=n_l+n_{\nu_l}=Y_{L_l}n_B$ (with $Y_{L_l}$ being the lepton
fractions) are also conserved.  In neutron star mergers one can adopt
the values $Y_{Le} =Y_{L\mu}=0.1$ for the neutrino-trapped
case~\cite{Baiotti2019}.

The $\beta$-equilibration rates of the processes~\eqref{eq:n_decay}
and \eqref{eq:l_capture} are given, respectively, by
\bea \label{eq:Gamma1_def}
\Gamma_{n\to pl\bar\nu} &=& \int d\Omega_p\sum \vert
{\cal M}_{\rm Urca}\vert^2 \bar{f}(k)\bar{f}(p)
\bar{f}(k') f(p') (2\pi)^4\delta^{(4)}(k+p+k'-p'),\\
\label{eq:Gamma2_def}
\Gamma_{pl\to n\nu} &=& \int d\Omega_p \sum \vert
{\cal M}_{\rm Urca}\vert^2  {f}(k) {f}(p)
\bar{f}(k') \bar{f}(p')(2\pi)^4\delta(k+p-k'-p').
\eea
where
\bea\label{eq:Omega}
\int d\Omega_p = \int\! \frac{d^3p}{(2\pi)^32p_0} \int\!
\frac{d^3p'}{(2\pi)^32p'_0} \int\! \frac{d^3k}{(2\pi)^32k_0}
\int\! \frac{d^3k'}{(2\pi)^32k'_0}
\eea
is the Lorentz-invariant momentum phase-space element, $f(p)$
is the Fermi {distribution} of particles, and $\bar{f}(p)=1-f(p)$.
The particles are assigned momenta as follows: $(l) \to k$,
$(\nu_l/\bar{\nu}_l) \to k'$, $(p)
\to p$, and $(n) \to p'$. Note that in neutrino-transparent
matter one should replace $\bar{f}(k')\to 1$ in these expressions.

The spin-averaged relativistic matrix element of the Urca processes
reads~\cite{Greiner2000gauge}
\bea\label{eq:matrix_el_full}
\sum \vert {\cal M}_{\rm Urca}\vert^2 &=& 32 G_F^2\cos^2
\theta_c \Big[(1+g_A)^2(k\cdot p) (k'\cdot p')\nonumber\\
&&+(1-g_A)^2(k\cdot p') (k'\cdot p)
+(g_A^2-1)m^{*2}(k\cdot k')\Big],
\eea
where $G_F=1.166\cdot 10 ^{-5}$ GeV$^{-2}$ is the Fermi coupling
constant, $\theta_c$ is the Cabibbo angle with $\cos\theta_c=0.974$,
$g_A=1.26$ is the axial-vector coupling constant, and $m^*$ is the
effective nucleon mass. We will keep only the first term of this
expression in the following as the second and the third terms are
negligible for $g_A$ values close to the vacuum value quoted
above. The twelve-dimensional phase-space integrals in
Equations~\eqref{eq:Gamma1_def} and \eqref{eq:Gamma2_def} can be reduced to
the following four-dimensional integrals which are then computed
numerically~\cite{Alford2021c}
\bea\label{eq:Gamma1p_final}
\Gamma_{n\to p l \bar\nu} (\mu_{\Delta_l})
&=& -\frac{{G}^2T^4}{(2\pi)^5}
\int_{-\infty}^\infty\!\!\! dy\,
\!\int_0^\infty\!\! dx\,
\left[(\mu_{\nu_l} +\mu_n^*+yT)^2
-m_n^{*2}-x^2T^2\right]\nonumber\\
&&\times \left[(\mu_l +\mu_p^* +\bar{y}_lT)^2
-m_l^2-m_p^{*2} -x^2T^2\right]\nonumber\\
&&\times \int_{m_l/T-\alpha_l}^{\alpha_p +\bar{y}_l}\!
dz\, \bar{f}(z){f}(z-\bar{y}_l)\,\theta_x\!
\int_{\alpha_{\nu_l}}^\infty\!
dz'\,f(z'+y)\bar{f}(z')\,\theta_y,\quad\\
\label{eq:Gamma2n_final}
\Gamma_{p l\to n\nu} (\mu_{\Delta_l})
&=& \frac{{G}^2T^4}{(2\pi)^5}\int_{-\infty}^\infty\!
dy\! \int_0^\infty\!
dx\, \left[(\mu_{\nu_l} +\mu_n^*+yT)^2
-m_n^{*2}-x^2T^2\right]\nonumber\\
&&\times \left[(\mu_l +\mu_p^* +\bar{y}_lT)^2
-m_l^2-m_p^{*2} -x^2T^2\right]\nonumber\\
&&\times \int_{m_l/T-\alpha_l}^{\alpha_p +\bar{y}_l}\!
dz\, f(z)f(\bar{y}_l-z)\,\theta_x\!
\int_{-\alpha_{\nu_l}}^{\alpha_n+y}\!
dz'\, {f}(z'-y)\bar{f}(z')\,\theta_z,
\eea
where $G=G_F \cos\theta_c(1+g_A)$, $m_l$ is the lepton mass,
$\alpha_l = \mu_l/T$, $\alpha_N= \mu_N^*/T$ for $N=\{n,p\}$ with
$\mu_N^*$ being the nucleon effective chemical potential, see
Section~\ref{sec:rates}. Here $\bar{y}_l=y+\mu_{\Delta_l}/T$ with
$\mu_{\Delta_l}=\mu_n+\mu_{\nu_l}-\mu_p-\mu_l$ and
$f(x)\equiv [1+e^x]^{-1}$. The $\theta$-functions in
Equations~\eqref{eq:Gamma1p_final} and~\eqref{eq:Gamma2n_final} imply

\bea\label{eq:thetax}
\theta_x &: &
(z_k-x)^2 \leq \left(z -\alpha_p
-\bar{y}_l\right)^2 -m_p^{*2}/T^2\leq (z_k+x)^2,\\
\label{eq:thetay}
\theta_y &: &
(z_k'-x)^2 \leq \left(z' +\alpha_n+ y\right)^2
-m_n^{*2}/T^2\leq (z_k'+x)^2,\\
\label{eq:thetaz}
\theta_z &: &
(z_k'-x)^2 \leq \left(z'-\alpha_n-y\right)^2
-m_n^{*2}/T^2\leq (z_k'+x)^2.
\eea

The integration variables $y$ and $x$ are the transferred energy
and momentum, respectively, normalized by the temperature; the
variables $z$ and $z'$ are the normalized-by-temperature lepton
and neutrino energies, respectively, computed from their chemical
potentials,  $z_k=\sqrt{(z+\alpha_l)^2-m_l^2/T^2}$ and
$z'_k=z'\mp \alpha_{\nu_l}$ are the normalized-by-temperature
momenta of the lepton and the antineutrino/neutrino, respectively.
In the case of\linebreak neutrino-trapped matter, the rates of the inverse
processes are obtained from\linebreak Equations~\eqref{eq:Gamma1p_final} and
\eqref{eq:Gamma2n_final} by replacing $f(p_i)\rightarrow \bar{f}(p_i)$
for all particles. In the case of $\nu$-transparent matter the
inverse processes are not allowed, and one should replace
$\mu_{\nu_l}=0$ and $\bar{f}(z')\to 1$ in the direct processes.

We will work in the low-temperature approximation where
beta equilibrium corresponds to $\mu_{\Delta_l}=0$. In the case
of  deviations from $\beta$-equilibrium, there is a net rate of proton
production/annihilation due to each of the processes~\eqref{eq:n_decay}
and \eqref{eq:l_capture}, which in the linear-response regime
$\mu_{\Delta_l} \ll T$ can be  written as
$\Gamma_{n\to p l \bar\nu}-\Gamma_{p l \bar\nu\to n} =
\lambda_{n\fromto p l \bar\nu}\,\mu_{\Delta_l}$, and
$\Gamma_{n\nu\to pl}-\Gamma_{p l\to n\nu} = \lambda_{p l\fromto n\nu}\,
\mu_{\Delta_l}$, with the coefficients $\lambda_{n\fromto p l \bar\nu}$
and $\lambda_{p l\fromto n\nu}$ given by~\cite{Alford2021c}
\bea \label{eq:lambda1}
\lambda_{n\fromto p l \bar\nu} &=&
\left(\frac{\partial\Gamma_{n\to p l \bar\nu}}
{\partial\mu_{\Delta_l}}-
\frac{\partial\Gamma_{p l \bar\nu\to n}}
{\partial\mu_{\Delta_l}}\right)\bigg\vert_{\mu_{\Delta_l}=0}
=\frac{\Gamma_{n\fromto p l \bar\nu}}{T},\\
\label{eq:lambda2}
\lambda_{p l\fromto n\nu} &= & \left(\frac{\partial\Gamma_{n\nu \to pl}}
{\partial\mu_{\Delta_l}}-
\frac{\partial\Gamma_{p l\to n\nu}}
{\partial\mu_{\Delta_l}}\right)\bigg\vert_{\mu_{\Delta_l}=0}
=\frac{\Gamma_{p l\fromto n\nu}}{T}.
\eea

Note that at temperatures $T\gtrsim 1\,\MeV$ and at densities
where direct Urca would be forbidden at $T=0$, the neutron decay
and lepton capture processes are Boltzmann-suppressed by different
factors, arising from their different phase spaces~\cite{Alford2018b}.
This means that the coefficients should be evaluated at a nonzero
$\mu_{\Delta_l} = \mu_{\Delta_l}^{\rm eq}$, but we work in the
approximation $\mu_{\Delta_l}^{\rm eq}=0$: this is discussed
in Section~\ref{sec:num_results}.

Similar to the Urca reaction rates, the lepton reaction
rates can be written in the following form
\bea \label{eq:Gamma_lep1}
\Gamma_{\mu\to e\bar{\nu}\nu}  =
\int d\Omega_k \sum \vert {\cal M}_{\rm lep}\vert^2
f(k_\mu) \bar{f}(k_e)\bar{f}(k_{\bar{\nu}_e}) \bar{f}(k_{\nu_\mu})
(2\pi)^4\delta^{(4)}(k_e+k_{\bar{\nu}_e}+k_{\nu_\mu}-k_\mu),\\
\label{eq:Gamma_lep2}
\Gamma_{\mu\nu\to e{\nu}} =
\int d\Omega_k \sum \vert {\cal M}_{\rm lep}\vert^2
f(k_\mu) {f}(k_{\nu_e})  \bar{f}(k_e)\bar{f}(k_{\nu_\mu})
(2\pi)^4\delta^{(4)}(k_e+k_{\nu_\mu}-k_{\nu_e}-k_\mu),\\
\label{eq:Gamma_lep3}
\Gamma_{\mu\bar{\nu}\to e\bar{\nu}}  =
\int d\Omega_k \sum \vert {\cal M}_{\rm lep}\vert^2
f(k_\mu){f}(k_{\bar{\nu}_\mu}) \bar{f}(k_e) \bar{f}(k_{\bar{\nu}_e})
(2\pi)^4\delta^{(4)}(k_e+k_{\bar{\nu}_e}-k_{\bar{\nu}_\mu}-k_\mu),
\eea
where $d\Omega_k$ is defined analogously to Equation~\eqref{eq:Omega}.
The spin-averaged relativistic matrix element of lepton reactions
reads~\cite{Guo:2020tgx}
\be\label{eq:matrix_lep}
\sum \vert {\cal M}_{\rm lep}\vert^2 = 128
G_F^2 \left(k_e\cdot k_{\nu_\mu/\bar{\nu}_\mu}\right)
\left(k_{\nu_e/\bar{\nu}_e}\cdot k_\mu\right).
\ee

The final expressions for the lepton reaction rates are
very similar to the Urca process rates \eqref{eq:Gamma1p_final} and
\eqref{eq:Gamma2n_final} and are given in Ref.~\cite{Alford2021c}.

\section{Bulk Viscosity of {\boldmath{{$npe \mu$}}} Matter}
\label{sec:bulk}

In this section, we briefly review the bulk viscosity of
relativistic $npe\mu$ matter arising from the Urca
processes~\eqref{eq:n_decay} and \eqref{eq:l_capture}.
For this, we consider small-amplitude density oscillations
with frequency $\omega$. Separating the oscillating parts
from the static equilibrium values of particle densities
we can write $n_j(t)=n_{j0}+\delta n_j(t)$, where $\delta
n_j(t)\sim e^{i\omega t}$ with $j=\{n,p,l,\nu_l\}$.
Oscillations drive the system out of chemical equilibrium
leading to nonzero chemical imbalances $\mu_{\Delta_l}=
\delta\mu_n+\delta\mu_{\nu_l}-\delta\mu_p-\delta\mu_l$,
which can be written as
\bea\label{eq:delta_mu_l}
\mu_{\Delta_l} = A_n \delta n_n +A_{\nu_e}
\delta n_{\nu_e} -A_p \delta n_p -A_l \delta n_l,
\eea
where the particle susceptibilites are defined as $A_n=A_{nn}-A_{pn}$,
$A_p=A_{pp}-A_{np}$, and $A_l=A_{ll}$, $A_{\nu_l}=A_{\nu_l \nu_l}$ with
\bea\label{eq:A_f}
A_{ij} = \frac{\partial \mu_i}{\partial n_j},
\eea
where the derivatives are computed in $\beta$-equilibrium state.

If the weak processes were switched off, then the number of
all particle species would conserve separately, which implies
\bea\label{eq:cont_j}
\frac{\partial}{\partial t} \delta {n}^0_j(t)+ \theta n_{j0} =0
\quad\Rightarrow\quad \delta {n}^0_j(t) = -\frac{\theta}{i\omega}\, n_{j0},
\eea
where $\theta=\partial_i v^i$ is the fluid expansion rate.
Once the weak reactions are switched on, there is a net production
of particles which should be included in the balance equations.
To linear order in chemical imbalances, these equations read
\bea\label{eq:cont_n_slow}
\frac{\partial}{\partial t}\delta n_n(t)
+ \theta  n_{n0} &=&
-\lambda_e\mu_{\Delta_e}(t)
-\lambda_{\mu} \mu_{\Delta_\mu}(t),\\
\label{eq:cont_p_slow}
\frac{\partial}{\partial t}\delta n_p(t)
+\theta n_{p 0} &=&
\lambda_e\mu_{\Delta_e}(t)
+\lambda_{\mu} \mu_{\Delta_\mu}(t),\\
\label{eq:cont_e_slow}
\frac{\partial}{\partial t}\delta n_e(t)
+\theta n_{e 0} &=&  \lambda_e\mu_{\Delta_e}(t)
+\lambda_L\mu_{\Delta}^L(t),\\
\label{eq:cont_mu_slow}
\frac{\partial}{\partial t}\delta n_\mu(t)
+ \theta n_{\mu 0}&=&
\lambda_{\mu}\mu_{\Delta_\mu}(t)
-\lambda_L\mu_{\Delta}^L(t),
\eea
where $\mu_{\Delta}^L\equiv \mu_\mu+\mu_{\nu_e}-\mu_e-\mu_{\nu_\mu}=
\mu_{\Delta_e}-\mu_{\Delta_\mu}$ is the chemical imbalance for leptons,
and $\lambda_l=\lambda_{n\fromto p
l \bar\nu}+\lambda_{p l\fromto n\nu}$.
The coefficient $\lambda_L$ is the purely leptonic analog to $\lambda_l$.

Solving the system of Equations~\eqref{eq:cont_n_slow}--\eqref{eq:cont_mu_slow}
is generally quite cumbersome. However, as shown in Section~\ref{sec:rates},
the lepton processes proceed typically much slower than the Urca
processes in both regimes of neutrino-transparent and neutrino-trapped
matter, \ie, $\lambda_{L}\ll \lambda_{l}$ (slow lepton-equilibration
limit). As a result, the terms $\propto\lambda_L$ can be dropped from
the balance Equations~\eqref{eq:cont_n_slow}--\eqref{eq:cont_mu_slow}.
In other words, the Urca-process-driven bulk viscosity can be
computed by assuming that the weak leptonic processes are frozen.

Substituting now Equation~\eqref{eq:delta_mu_l} in Equations~\eqref{eq:cont_n_slow}
and \eqref{eq:cont_e_slow} and putting $\lambda_L=0$ we find
\bea
\label{eq:delta_ne}
i\omega\delta n_n &=&
-n_{n0}\theta -(\lambda_e+\lambda_{\mu}) A_n\delta n_n
+(\lambda_e+\lambda_{\mu}) A_p\delta n_p +\lambda_e A_e\delta n_e\nonumber\\
&+&\lambda_{\mu} A_\mu\delta n_\mu-\lambda_e A_{\nu_e} \delta n_{\nu_e}
-\lambda_{\mu} A_{\nu_\mu} \delta n_{\nu_\mu},\\
\label{eq:delta_nn}
i\omega\delta n_e &=& -n_{e0}\theta +\lambda_e A_n\delta n_n
- \lambda_e A_p\delta n_p-\lambda_e A_e\delta n_e+\lambda_e
A_{\nu_e} \delta n_{\nu_e}.
\eea

Using the relations $\delta n_p+\delta n_n=\delta n_B$,
$\delta n_e+\delta n_\mu=\delta n_p$, $\delta n_{L_e}
=\delta n_e+\delta n_{\nu_e}$, and $\delta n_{L_\mu}=
\delta n_\mu+\delta n_{\nu_\mu}$ and solving the coupled
Equations~\eqref{eq:delta_ne} and \eqref{eq:delta_nn} we
find ($\lambda\equiv \lambda_e +\lambda_{\mu}$)

\bea\label{eq:delta_nn1}
D\delta n_n
&=&-\frac{\theta}{i\omega}\bigg\{i\omega \left[n_{n0}(i\omega+\lambda_e A_e
+\lambda_e A_{\nu_e}) +n_{e0}(\lambda_e A_e+\lambda_e A_{\nu_e}-
\lambda_{\mu} A_\mu -\lambda_{\mu} A_{\nu_\mu})\right]\nonumber\\
&+&\left[i\omega(\lambda A_p+ \lambda_{\mu} A_\mu+\lambda_{\mu} A_{\nu_\mu})
+\lambda_e \lambda_{\mu} (( A_1- A_n)( A_2-A_n)- A_p^2)\right]n_{B0}\nonumber\\
&-& \lambda_e A_{\nu_e}\left[i\omega +\lambda_{\mu} (A_\mu+ A_{\nu_\mu})
\right] n_{L_e0} -\lambda_{\mu} A_{\nu_\mu}\left[i\omega+\lambda_e
(A_e+ A_{\nu_e})\right] n_{L_\mu 0}\bigg\},\\
\label{eq:delta_ne1}
D\delta n_e
&=&-\frac{\theta}{i\omega}\bigg\{i\omega n_{e0}
\Big[i\omega +\lambda_{\mu} A_2+\lambda_e (A_n+A_p)\Big]\nonumber\\
&-& \lambda_e  n_{B0} \Big[A_p
(i\omega+\lambda_{\mu} A_2)-\lambda_{\mu}(A_n+A_p)(A_2-A_n)\Big]\nonumber\\
&+&\lambda_e  n_{L_e0}A_{\nu_e}
(i\omega+\lambda_{\mu} A_2)
+\lambda_e (A_n+A_p) i\omega n_{n0}  \nonumber\\
&-&\lambda_e \lambda_{\mu}(A_n+A_p) A_{\nu_\mu} n_{L_\mu 0}\bigg\},
\eea
where we used the baryon and lepton number conservation
$\delta n_B =-n_{B0}(\theta/i\omega)$ and $\delta n_{L_l}
=-n_{L_l0}(\theta/i\omega)$, and defined
\bea\label{eq:D}
D
=(i\omega+\lambda_e A_1)(i\omega+\lambda_{\mu} A_2)
-\lambda_e \lambda_{\mu}(A_n+A_p)^2
\eea
with
\bea
\label{eq:def_A1}
A_1 &=& A_n+A_p+A_{e}+A_{\nu_e},\\
\label{eq:def_A2}
A_2 &=& A_n+A_p+A_{\mu}+A_{\nu_\mu}.
\eea

In order to find the bulk viscosity we still need to separate
the instantaneous equilibrium parts of particle densities from
perturbations~\eqref{eq:delta_nn1} and \eqref{eq:delta_ne1}.
Equilibrium shifts can be obtained from Equations~\eqref{eq:delta_nn1}
and \eqref{eq:delta_ne1} either in the limit of $\lambda_{l}\to
\infty$ (fast equilibration), or in the limit of $\lambda_{l}\to 0$
(slow equilibration). Both choices lead us to the same result for
the bulk viscosity as the latter vanishes in both limits of
fast or slow equilibration.
Subtracting thus the local quasi-equilibrium shifts $\delta n_j^{0}$
from Equations~\eqref{eq:delta_nn1} and \eqref{eq:delta_ne1} we find the
required nonequilibrium parts $\delta n'_j = \delta n_j-\delta n^{0}_j$.
After this the nonequilibrium part of the pressure,
referred to as bulk viscous pressure, will be given by
\bea\label{eq:Pi}
\Pi =\sum_j
c_j\delta n'_j,
\eea
with
\bea\label{eq:c_j_def}
c_j \equiv  \frac{\partial p}{\partial n_j}=\sum_i n_{i0}
\frac{\partial \mu_i}{\partial n_j}=\sum_i n_{i0}A_{ij}.
\eea

Here we used the Gibbs--Duhem relation $dp=sdT+\sum_i n_i d \mu_i$,
and recalled the definitions~\eqref{eq:A_f}.
The bulk viscous pressure then reads
\bea\label{eq:Pi_slow1}
\Pi = \frac{\theta}{i\omega}
\frac{i\omega (\lambda_e C_1^2+\lambda_{\mu} C_2^2)
+\lambda_e \lambda_{\mu} \big[A_1 C_2^2 +A_2 C_1^2-2(A_n+A_p)C_1C_2\big]}
{(i\omega+\lambda_e A_1)(i\omega+\lambda_{\mu} A_2)
-\lambda_e \lambda_{\mu}(A_n+A_p)^2},
\eea
where we defined
\bea
\label{eq:def_C1}
c_n-c_p-c_e +c_{\nu_e}=n_{n0}A_n-n_{p0}A_{p}-
n_{e0}A_e+n_{\nu_e0}A_{\nu_e}\equiv C_1,\\
\label{eq:def_C2}
c_n-c_p-c_\mu +c_{\nu_\mu}=n_{n0}A_n-n_{p0}A_{p}-
n_{\mu0}A_\mu+n_{\nu_\mu0}A_{\nu_\mu}\equiv C_2.
\eea
Extracting the real part of Equation~\eqref{eq:Pi_slow1} and recalling the
definition of the bulk viscosity ${\rm Re}\Pi=-\zeta \theta$ we find

\begin{adjustwidth}{-\extralength}{0cm}
\bea\label{eq:zeta_slow}
\zeta = \frac{\lambda_e \lambda_{\mu}\Big\{\lambda_e
\left[(A_n+A_p) C_1- A_1 C_2\right]^2+\lambda_{\mu}
\left[(A_n+A_p) C_2- A_2 C_1\right]^2 \Big\}+
\omega^2(\lambda_e C_1^2+\lambda_{\mu} C_2^2)}
{\Big\{\lambda_e \lambda_{\mu}\left[A_1A_2-(A_n+A_p)^2\right]-\omega^2\Big\}^2
+\omega^2(\lambda_e A_1+\lambda_{\mu} A_2)^2}.
\eea
\end{adjustwidth}
If we neglect the muonic contribution then we arrive at
\bea\label{eq:zeta_slow1}
\zeta_e = \frac{C_1^2}{A_1}\frac{\gamma_e}
{\omega^2+\gamma_e^2},
\eea
with $\gamma_e=\lambda_e A_1$, which coincides with
the result of Ref.~\cite{Alford2019b}.

In the limit of high frequencies $\omega\gg \lambda_l A_i$
we find from Equation~\eqref{eq:zeta_slow}
\bea\label{eq:zeta_slow2}
\zeta = \frac{\lambda_e C_1^2+\lambda_{\mu} C_2^2}
{\omega^2}=\zeta_e+\zeta_\mu,
\eea
where $\zeta_e$ and $\zeta_\mu$ are the partial bulk viscosities by
electronic and muonic Urca processes, respectively~\cite{Haensel2000}.

In the opposite limit of low frequencies we find
\bea\label{eq:zeta_slow3}
\zeta
=\frac{\lambda_e(C_1- a_1 C_2)^2
+\lambda_{\mu} (C_2- a_2 C_1)^2}
{\lambda_e \lambda_{\mu}(A_n+A_p)^2 (a_1 a_2-1)^2},
\eea
with $a_1=A_1/(A_n+A_p)$ and $a_2=A_2/(A_n+A_p)$.

\section{Numerical Results}
\label{sec:num_results}



The numerical calculations are performed within the framework
of covariant density functional approach to the nuclear matter
with density-dependent nucleon--meson couplings.
The Lagrangian density reads
\bea\label{eq:lagrangian}
{\cal L} & = &
\sum_N\bar\psi_N\bigg[\gamma^\mu \left(i\partial_\mu-g_{\omega}
\omega_\mu - \half g_{\rho }\vectau\cdot\vecrho_\mu\right)
- m^*_N\bigg]\psi_N + \sum_{\lambda}\bar\psi_\lambda
(i\gamma^\mu\partial_\mu - m_\lambda)\psi_\lambda,\\
\nonumber & + & \half \partial^\mu\sigma\partial_\mu\sigma-
\half m_\sigma^2\sigma^2 
- \frac{1}{4}\omega^{\mu\nu}
\omega_{\mu\nu} + \half m_\omega^2\omega^\mu\omega_\mu -
\frac{1}{4}\vecrho^{\mu\nu}\vecrho_{\mu\nu} + \half
m_\rho^2\vecrho^\mu\cdot\vecrho_\mu,
\eea
where $N$ sums over nucleons, $\psi_N$ are the nucleonic Dirac
fields, $m_N^*=m_N - g_{\sigma}\sigma$ are the nucleon effective
masses, with $m_N$ being the nucleon mass in vacuum. Next,
$\sigma,\omega_\mu$, and $\vecrho_\mu$ are the scalar-isoscalar,
vector-isoscalar, and vector-isovector meson fields, respectively;
$\omega_{\mu\nu}=\partial_\mu\omega_\nu-\partial_\nu\omega_\mu$
and $\vecrho_{\mu\nu}=\partial_\mu \vecrho_{\nu}-\partial_\nu
\vecrho_{\mu}$ are the field strength tensors of vector mesons;
$m_{i}$ are the meson masses and $g_{i}$ are the baryon-meson
couplings with $i=\sigma,\omega,\rho$. Finally, $\psi_\lambda$ are the
leptonic free Dirac fields with masses $m_\lambda$. Below we will
adopt the DDME2 parametrization of the couplings
$g_{i}$~\cite{Lalazissis2005} with the numerical implementation given
in Ref.~\cite{Colucci2013}.

\begin{figure}[H]
\includegraphics[width=0.45\columnwidth, keepaspectratio]{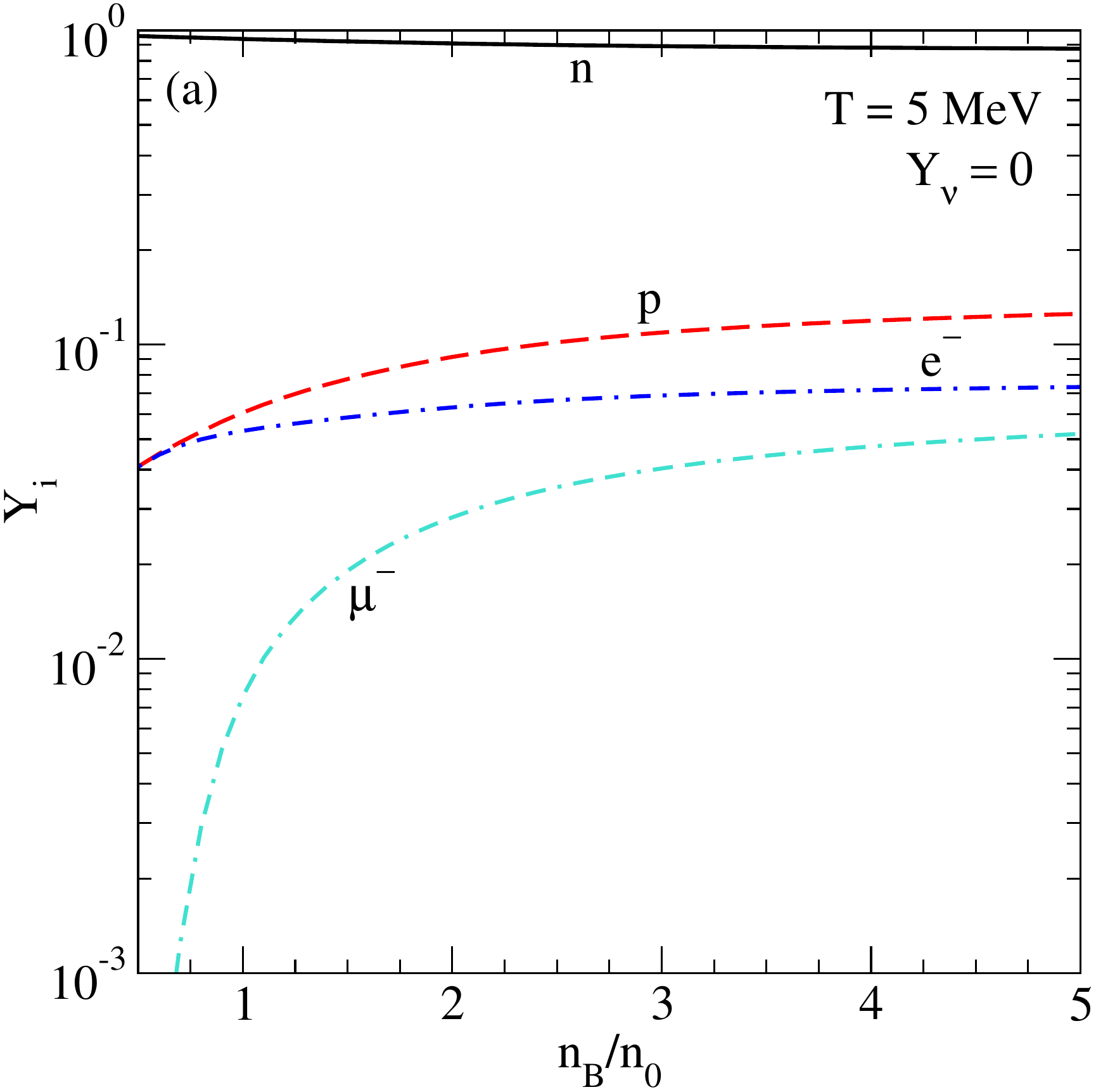}
\hspace{0.5cm}
\includegraphics[width=0.45\columnwidth, keepaspectratio]{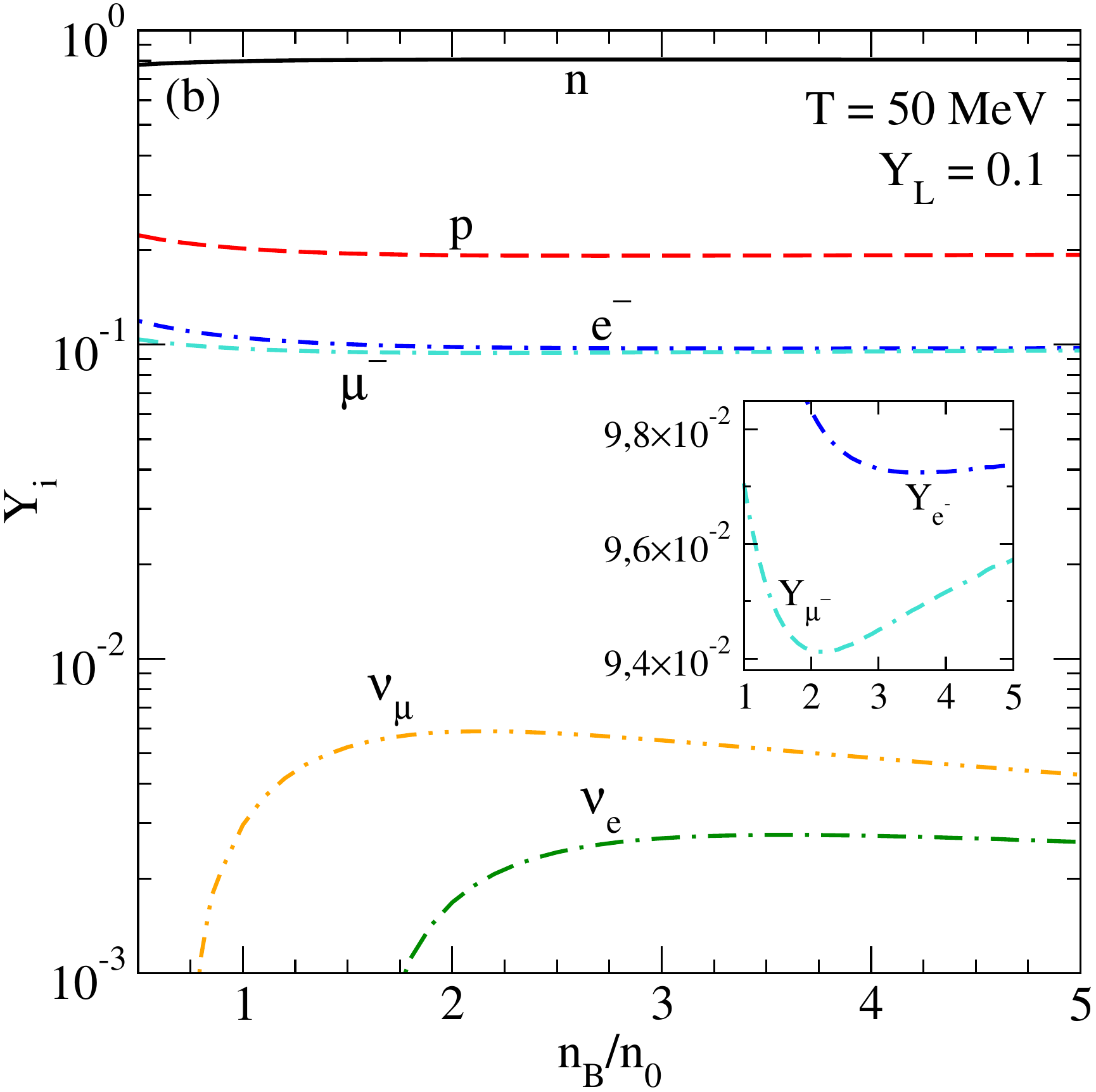}
\caption{{Composition} 
of neutron star merger matter in the DDME2 model, in neutrino-transparent
regime with $T=5$~MeV (\textbf{a}) and neutrino-trapped regime with $T=50$~MeV (\textbf{b}). The inset shows the
minima in the electron and $\mu$-on fractions, at which the subsystems of electrons and muons
are scale-invariant, and the corresponding partial bulk viscosities vanish.}
\label{fig:fractions}

\end{figure}

The composition of beta-equilibrated $npe\mu$ matter in two regimes of low and high temperatures is shown in Figure~\ref{fig:fractions}. The proton fraction in the neutrino-transparent matter remains below the threshold value required for the direct Urca processes to operate in the low-temperature regime in the whole density range considered here ($0.5n_0\leq n_B\leq 5n_0$). The threshold values of the proton fraction for electronic and muonic Urca processes are $Y_p\approx 13\%$ and $Y_p\approx 16\%$, respectively, whereas the proton fraction remains below  12.5\% up to the density $n_B = 5n_0$.
Note that in this case at
very low densities the net neutrino densities become negative,
indicating that the matter contains more antineutrinos than neutrinos
in that regime.

\subsection{Equilibration Rates of Weak Processes}
\label{sec:rates}



\begin{figure}[H]
\includegraphics[width=0.45\columnwidth,keepaspectratio]{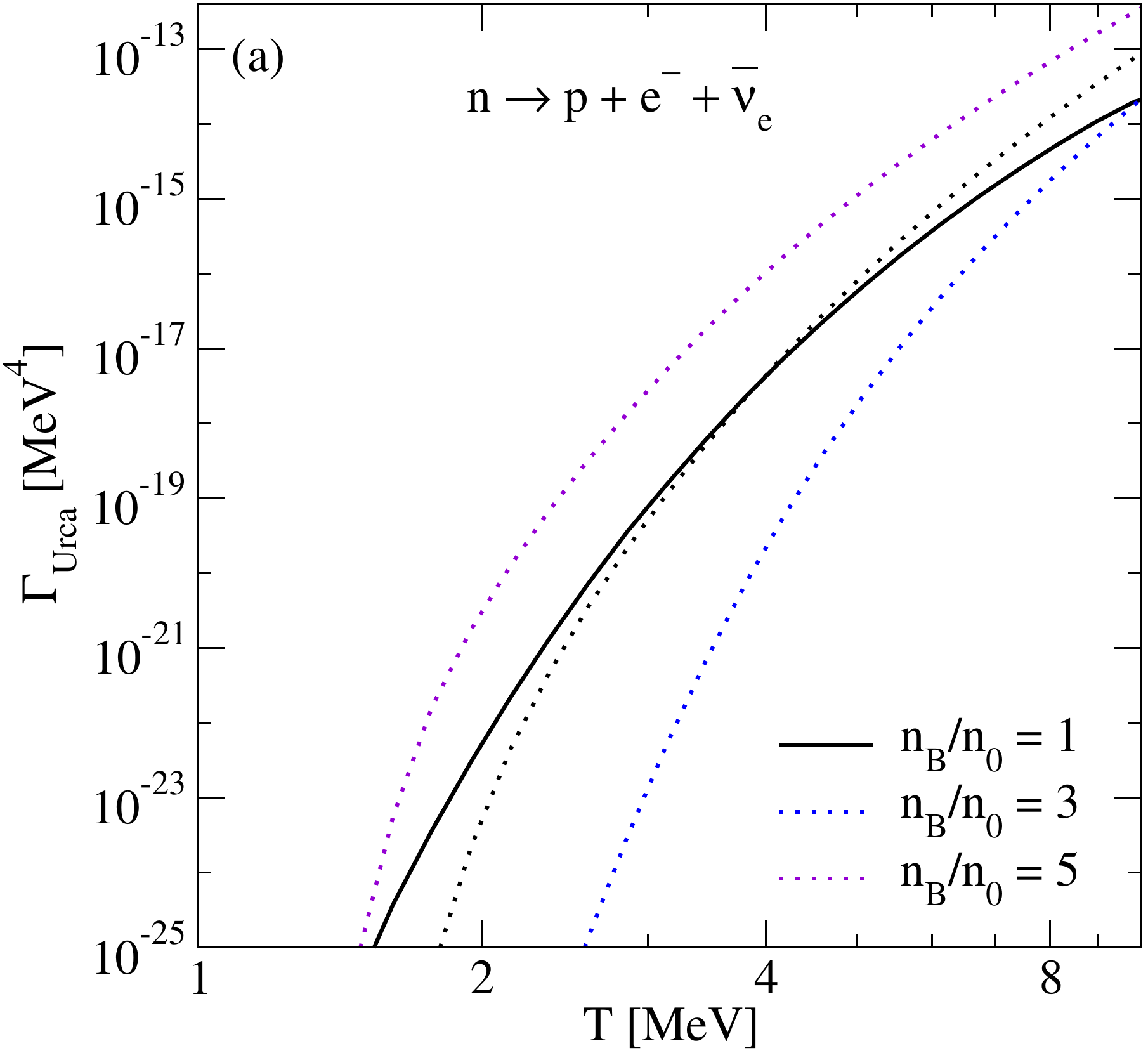}
\hspace{0.5cm}
\includegraphics[width=0.45\columnwidth,keepaspectratio]{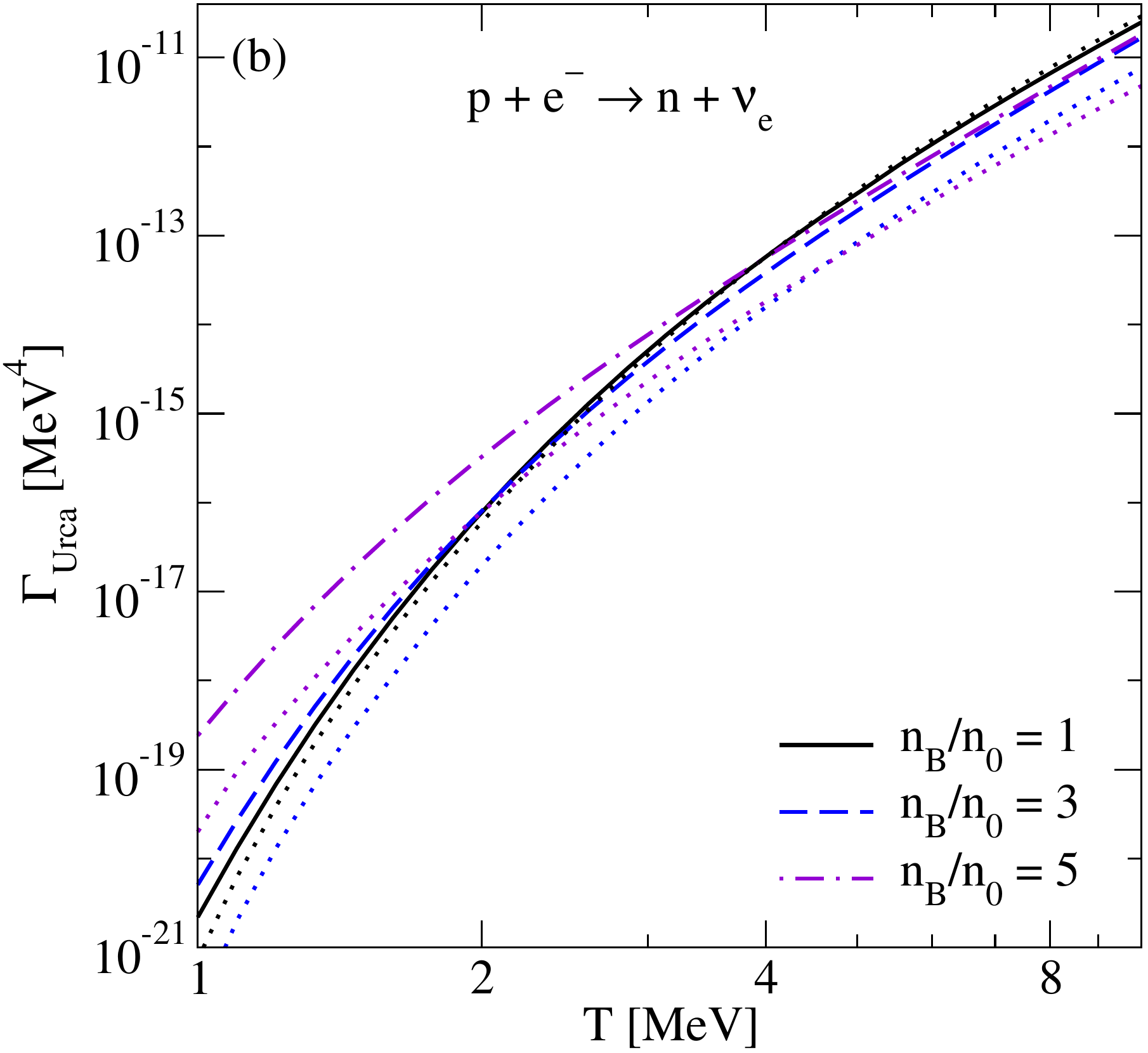}
\caption{The rates of neutron $e$-decay (\textbf{a}) and electron capture (\textbf{b}) processes
as functions of the temperature for various densities for neutrino-transparent
matter. The dotted lines show the rates of the same processes computed in
{Ref.}
~\cite{Alford2019b} within the approximation of nonrelativistic nucleons.}
\label{fig:Gamma12_e_trans}
\end{figure}

The electron-producing neutron decay
and electron capture rates for
neutrino-{\linebreak}transparent matter are shown in
Figure~\ref{fig:Gamma12_e_trans} as functions of the temperature.
The equilibration rates rapidly increase with increasing temperature as a
result of the fast opening of the scattering phase space. We see also that
the neutron decay rate is suppressed as compared to the electron
capture rate at least by three orders of magnitude, and is
exponentially damped at low temperatures and high densities because of
diminished scattering phase space (there are no curves corresponding
to $n_B=3n_0$ and $n_B=5n_0$ in panel (a) as the rate is highly damped
in these cases).
Similar behavior for the neutron decay rate was found
also for other EoS models in Ref.~\cite{Alford2021b}. As a result,
under the condition
$\mu_n=\mu_p+\mu_e$ the neutron decay and electron capture rates do
not balance each other, which implies that the matter is out of
$\beta$-equilibrium. As noted in the discussion of Equations~\eqref{eq:lambda1}
and \eqref{eq:lambda2}, in principle this shows the need for
a nonzero isospin chemical potential.
However, as the main  focus of this work is to study how the muonic reactions contribute to the bulk viscosity of $npe\mu$ matter, below we will neglect that finite temperature correction, given that it would not change the value of bulk viscosity at the maximum, and  (because the rates are so sensitive to temperature) would only shift the temperature at which that maximum is attained by about 1 MeV.

We show also the
electron-capture rates in Figure~\ref{fig:Gamma12_e_trans} which were
computed in Ref.~\cite{Alford2019b} in the approximation of
nonrelativistic nucleons by dotted lines. We see that the
nonrelativistic approximation underestimates the exact electron
capture rates by up to an order of magnitude (at $n_B=5n_0$).
This is not the case for the neutron decay process which shows finite
nonrelativistic rates also at high densities where the exact
relativistic calculations predict their strong suppression by an
exponential (Boltzmann-type) factor.

\begin{figure}[H]
\includegraphics[width=0.45\columnwidth,keepaspectratio]{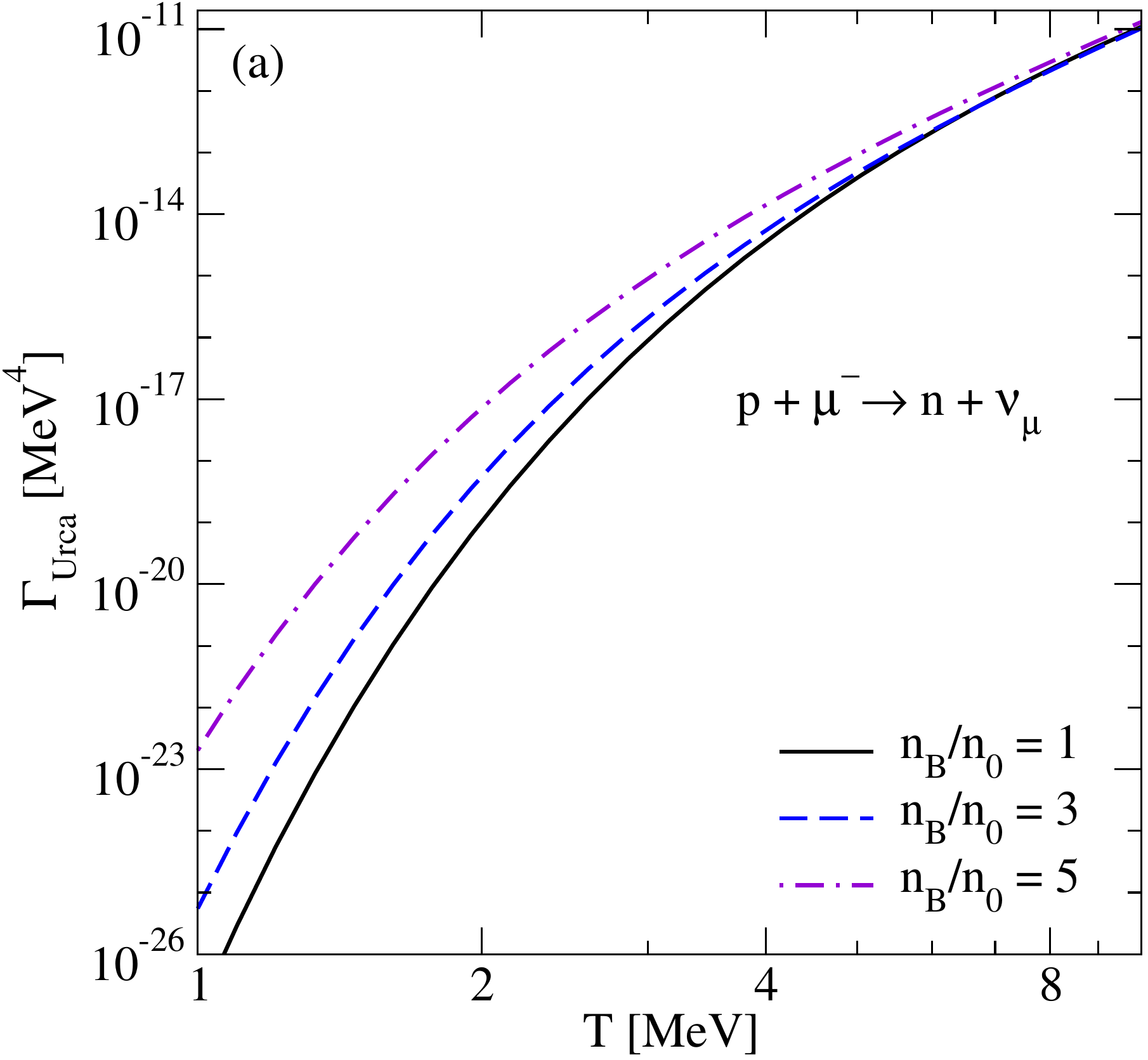}
\hspace{0.5cm}
\includegraphics[width=0.45\columnwidth,keepaspectratio]{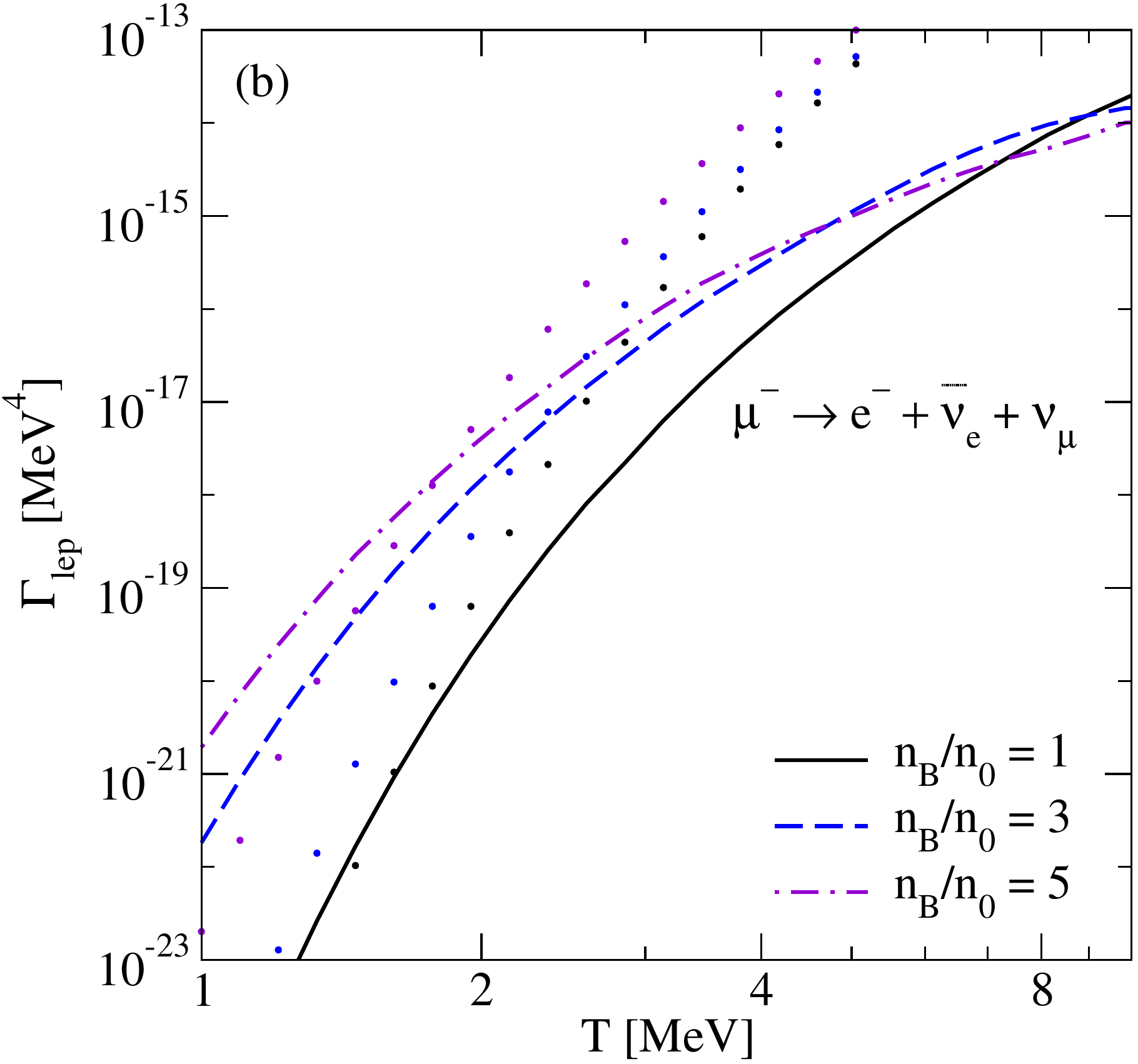}
\caption{The rates of muon capture (\textbf{a}) and muon decay (\textbf{b}) processes as
functions of the temperature for various densities for neutrino-transparent
matter. The neutron $\mu$-decay rate is strongly damped as compared to the
muon capture rate. The dotted lines in panel (\textbf{b}) show the muon capture rates.}
\label{fig:Gamma_mu_trans}
\end{figure}

Panel (a) of Figure~\ref{fig:Gamma_mu_trans} shows the muon capture rates,
the general behavior of which is similar to the electron capture rates.
Quantitatively, the muon capture rate is much smaller at low temperatures
and becomes comparable to the electron capture rate above $T\geq 5$ MeV.
The rate of the neutron $\mu$-decay is
always smaller than those of other processes that affect muon
density by at least three orders of magnitude
over the density and temperature range considered here and
is not shown.

Panel (b) of Figure~\ref{fig:Gamma_mu_trans} shows the rate of the muon
decay~\eqref{eq:reaction_L1}. The muon decay process has the same kinematics
as the neutron decay. As a result, the temperature dependence of
the muon decay rate is qualitatively very similar to that of neutron decay if it is finite. To compare the Urca and leptonic reaction rates we show in
Figure~\ref{fig:Gamma_mu_trans}b the muon capture rates by dotted lines
(electron capture rates are much larger than the muon decay rates and are
not shown here). We see, that, typically, the Urca reaction rates are much
larger than the leptonic reaction rates, the only exception being the range
of very low temperatures $T\lesssim 2$ MeV,
where both processes involving  muons are much slower than
electron capture process.  In this narrow range of
temperatures, the muonic contribution to the bulk viscosity can be neglected,
whereas at higher temperatures both electronic and muonic Urca processes
should be accounted for with leptonic reactions assumed to be frozen, as
discussed in Section~\ref{sec:bulk}.



\begin{figure}[H]
\includegraphics[width=0.45\columnwidth,keepaspectratio]{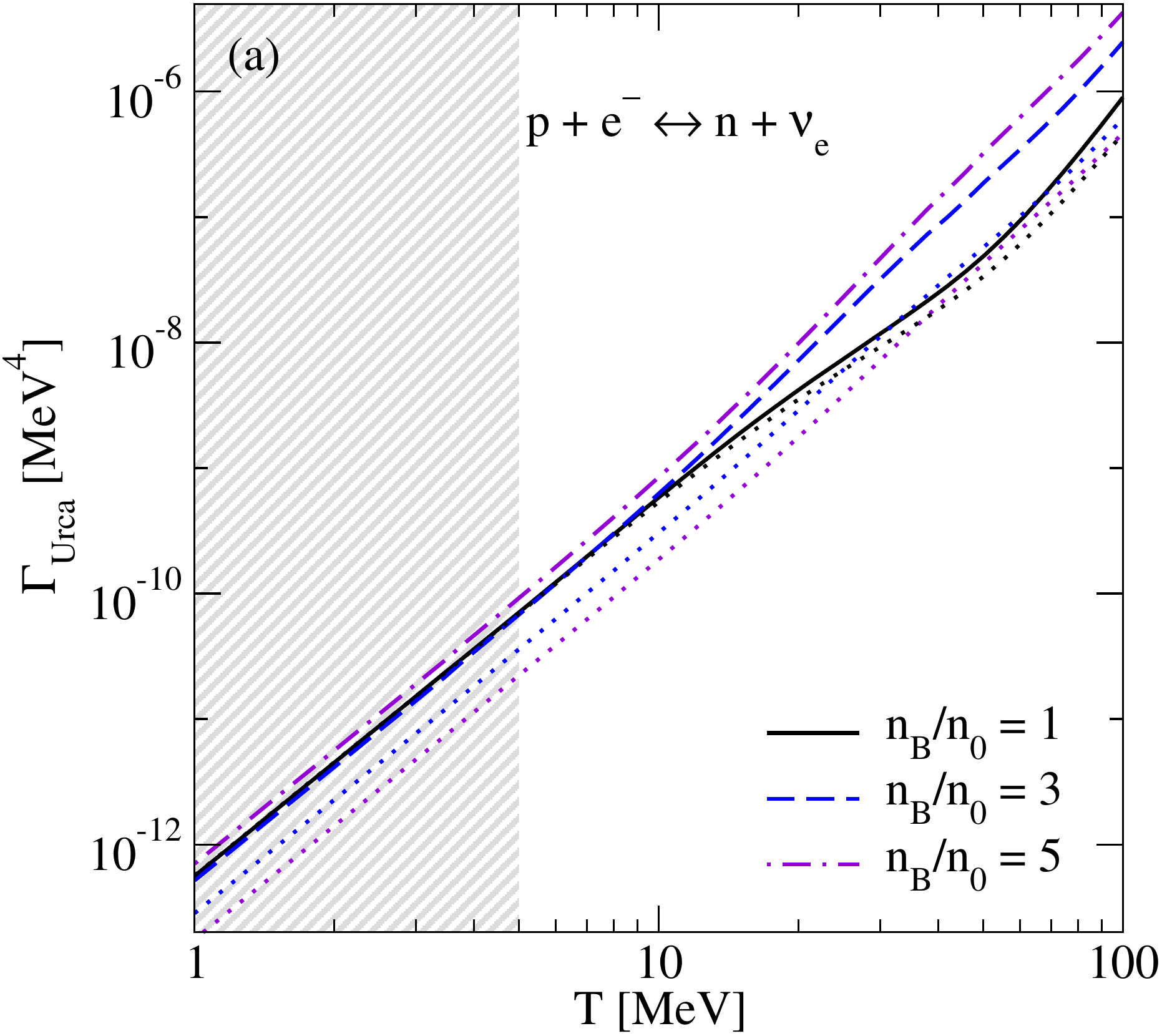}
\hspace{0.5cm}
\includegraphics[width=0.45\columnwidth,keepaspectratio]{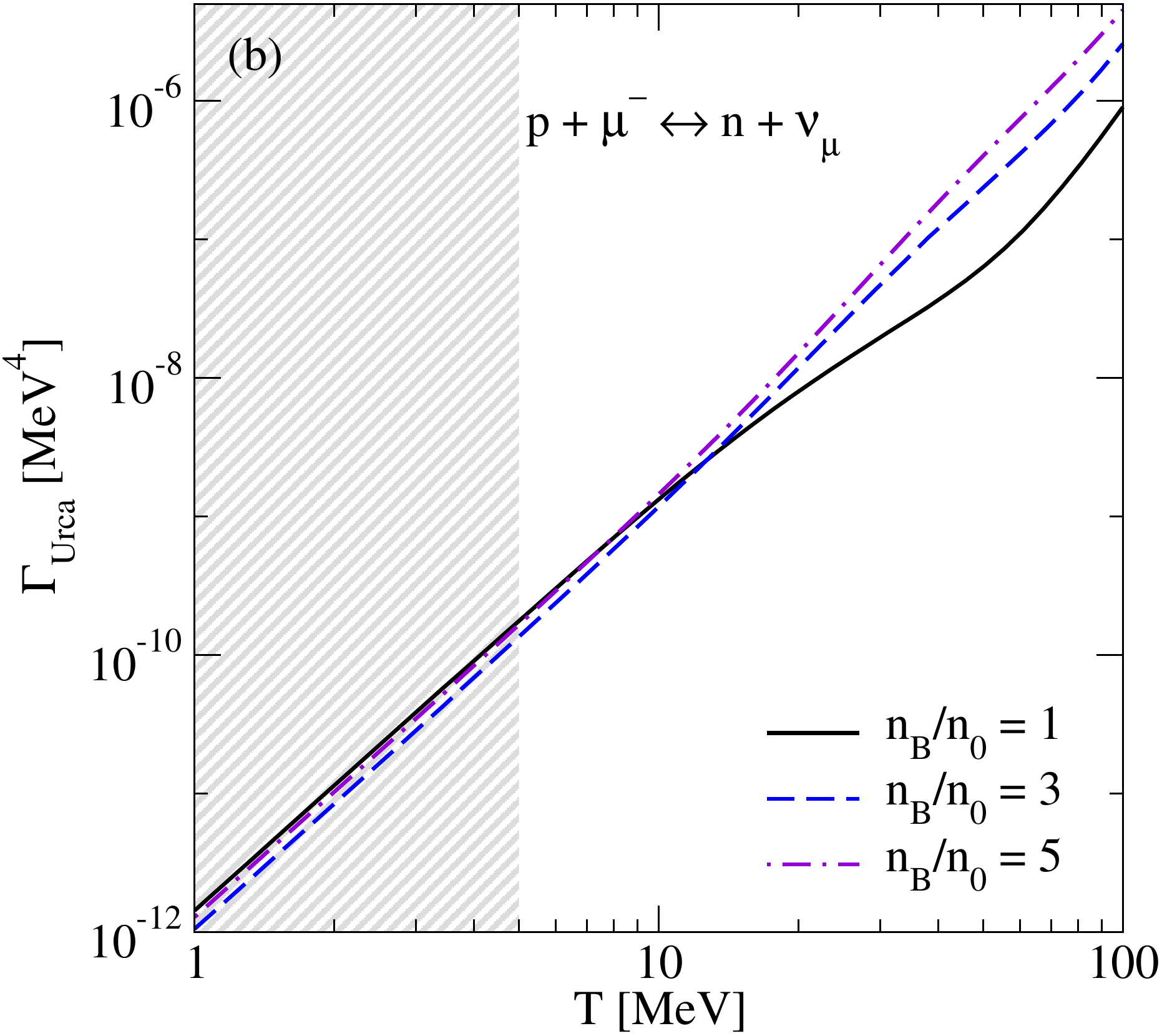}
\caption{The electron (\textbf{a}) and muon (\textbf{b}) capture rates as functions of
the temperature for various densities for the neutrino-trapped
matter. The neutron decay rates are negligible compared to the
lepton capture rates in the whole temperature-density range. {
The shaded areas show the neutrino transparent region of
temperatures $T \le 5$ MeV.  } The dotted lines in panel (\textbf{a}) show
the nonrelativistic approximation to the electron capture rates as
computed in {Ref.} 
~\cite{Alford2019b}.}
\label{fig:Gamma_Urca_trap}
\end{figure}

Figure~\ref{fig:Gamma_Urca_trap} shows the rates of the electron (a)
and muon (b) capture processes in the neutrino-trapped regime. At
moderate temperatures $T\leq 10$~MeV, the lepton capture rates follow
their low-temperature scaling $\propto T^3$~\cite{Alford2021c}. The
electron and muon capture rates are very similar both qualitatively
and quantitatively. Panel (a) shows also the electron-capture rates of
nuclear matter in the approximation of nonrelativistic
nucleons~\cite{Alford2019b}.  As in the neutrino-transparent case, the
non-relativistic approximation underestimates the exact equilibration
rates also in the neutrino-trapped regime by a factor that rises with
the density from 1 to 10. As for the neutron decay
processes~\eqref{eq:n_decay}, their rates are many orders of magnitude
smaller than the lepton capture rates as the formers involve
antineutrinos instead of neutrinos. A detailed discussion on the
relative importance of the neutron decay and lepton capture rates can
be found in Ref.~\cite{Alford2021c}.

\begin{figure}[H]
\includegraphics[width=0.45\columnwidth,keepaspectratio]{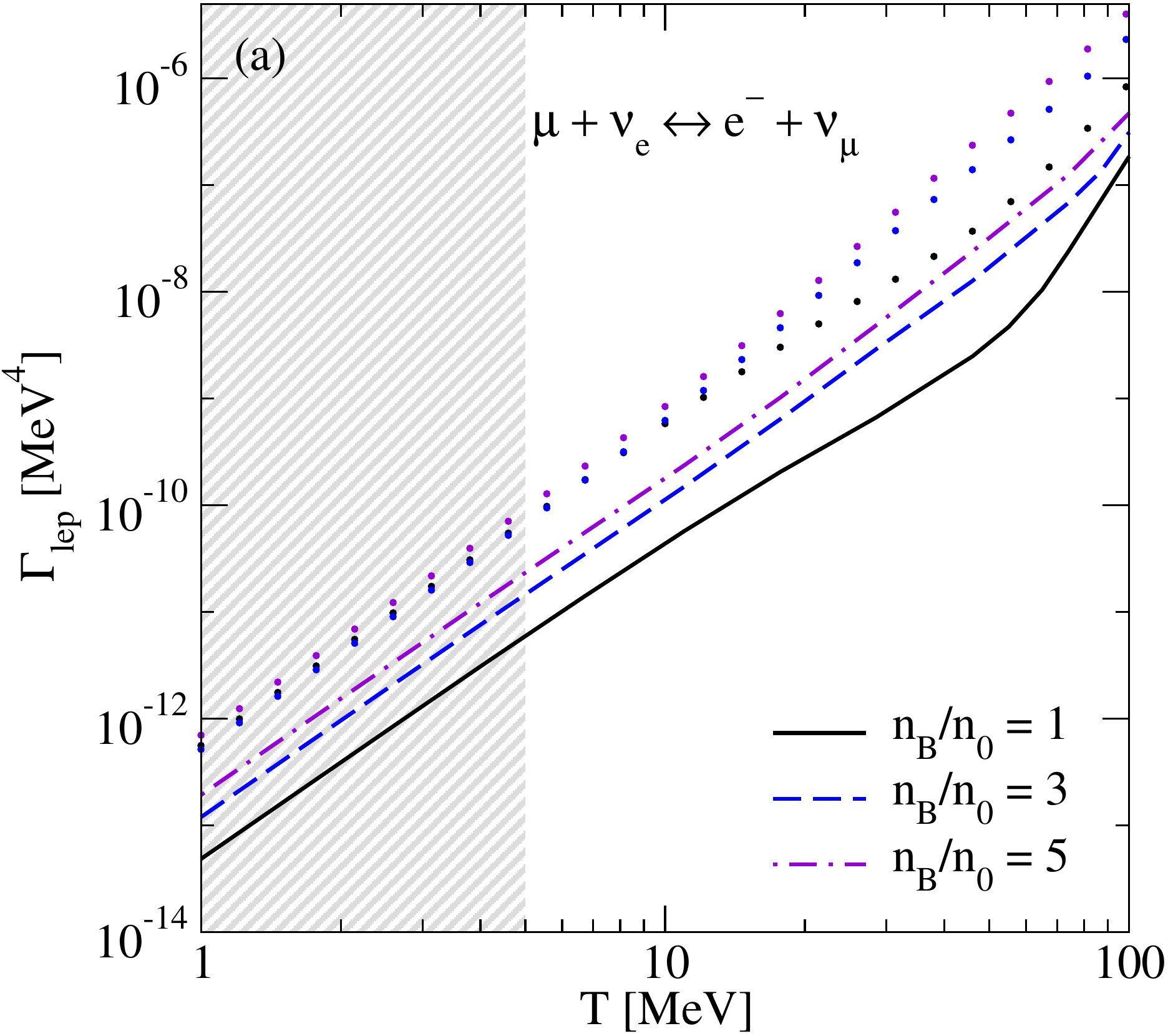}
\hspace{0.5cm}
\includegraphics[width=0.45\columnwidth,keepaspectratio]{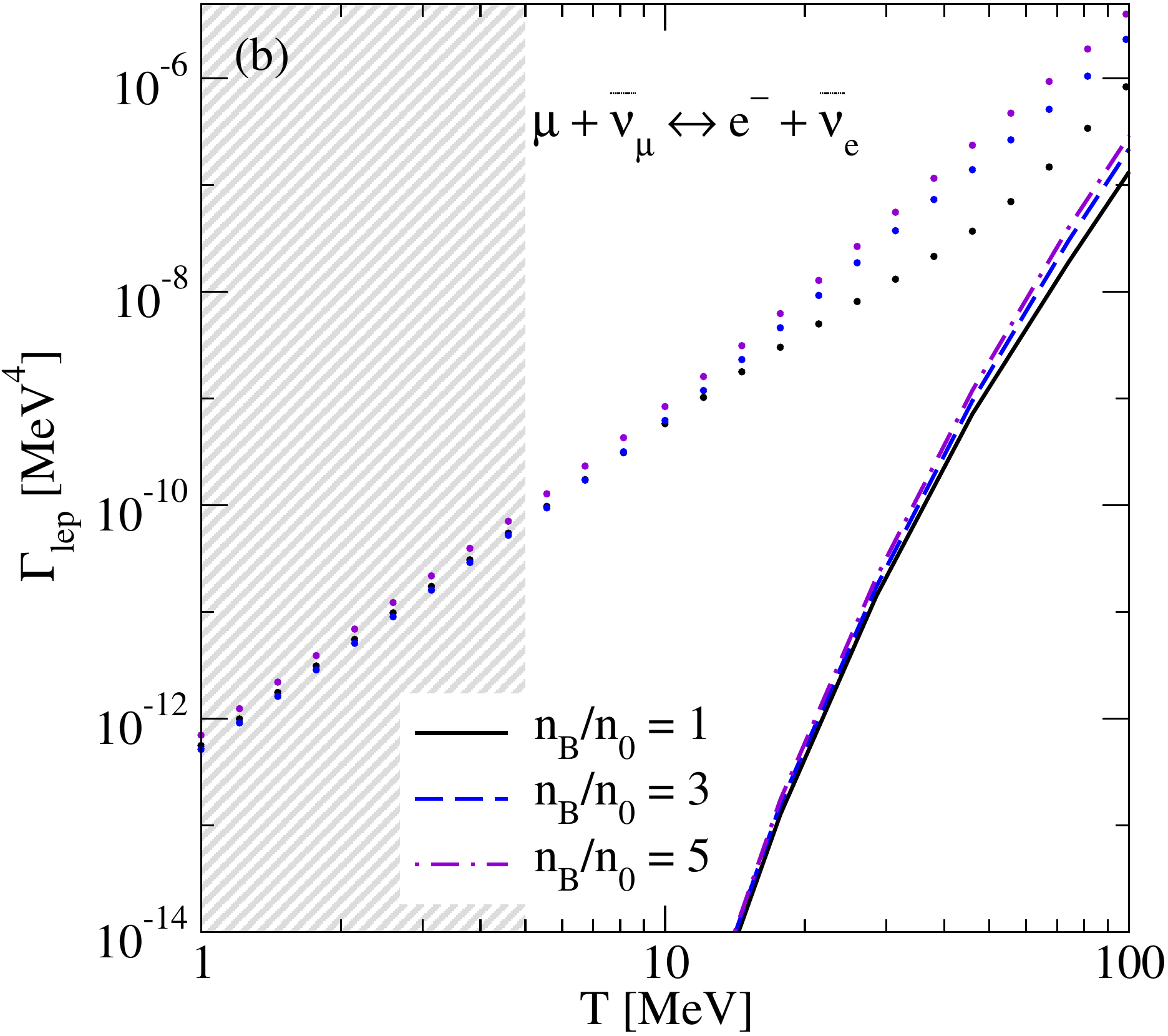}
\caption{Rates of the neutrino (\textbf{a}) and antineutrino (\textbf{b}) absorption
processes as functions of the temperature for different values of
the density for the neutrino-trapped matter. The electron (Urca) capture
rates are shown by dotted lines for comparison; the muon capture rates
are slightly higher than the electron capture rates and are not shown. {The shaded areas show the
neutrino transparent region of temperatures $T \le 5$ MeV.}
}
\label{fig:Gamma_lep_trap}
\end{figure}

Figure~\ref{fig:Gamma_lep_trap} shows the rates of neutrino (a) and
antineutrino (b) absorption processes. The neutrino absorption rates
show similar to the lepton capture rates
behavior (shown by dotted lines) but are smaller on average by
an order of magnitude. As expected, the antineutrino absorption
rates are much smaller than the neutrino absorption rates. The
muon decay rate is even smaller than the antineutrino absorption
process because of the very small scattering phase space. Thus,
we conclude that the leptonic processes are always much slower
than the Urca processes also in the neutrino-trapped matter.

\subsection{Susceptibilities and Urca Relaxation Rates}
\label{sec:susc}

We have extended our work on the bulk viscosities in the
isothermal regime to the case of isentropic matter. Here among other
things we compare the isothermal and isentropic results leaving
the  detailed discussion of the latter to a future
work~\cite{Alford2022preprint}.

\begin{figure}[H]
\includegraphics[width=0.45\columnwidth,keepaspectratio]{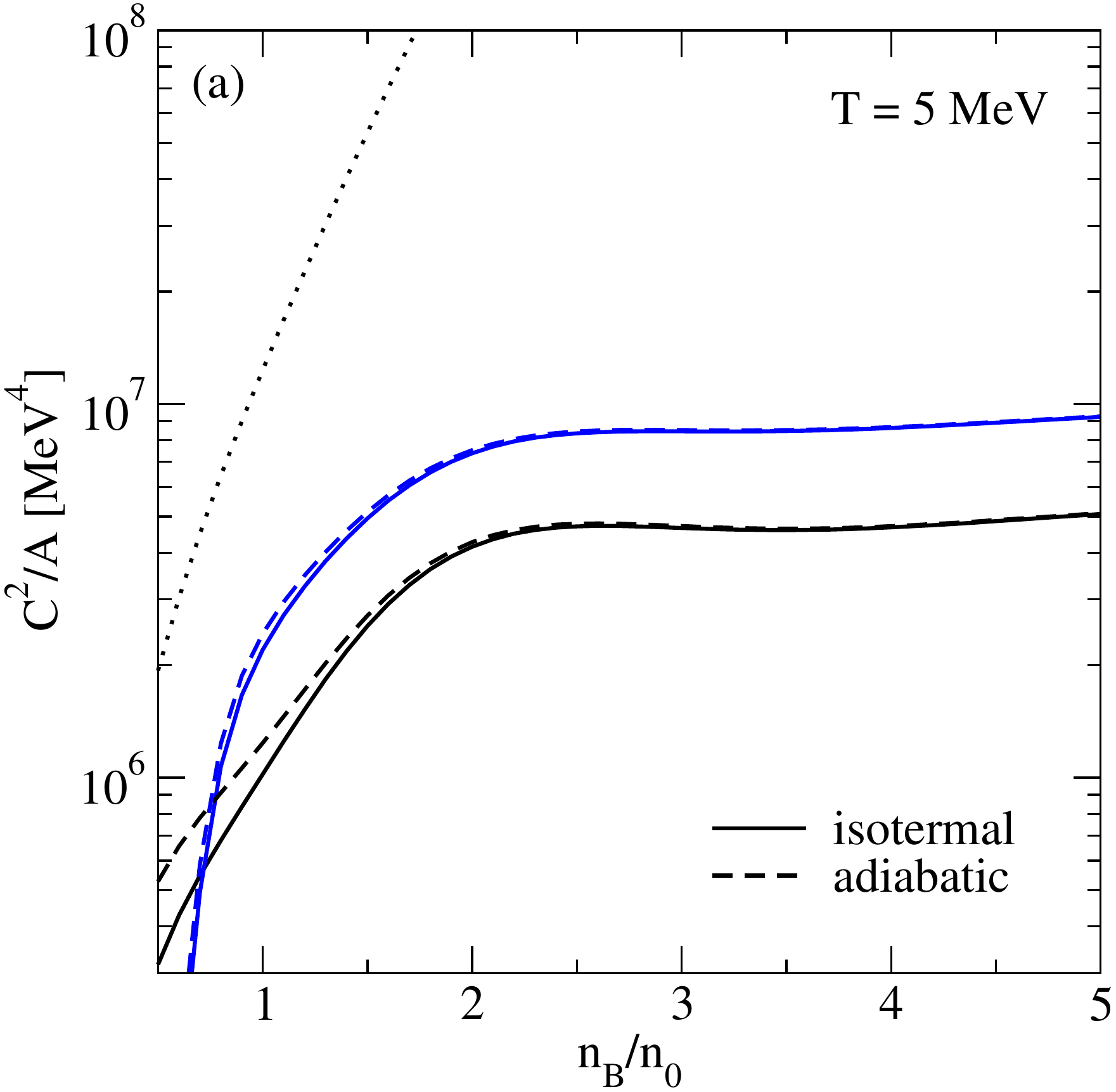}
\hspace{0.5cm}
\includegraphics[width=0.45\columnwidth,keepaspectratio]{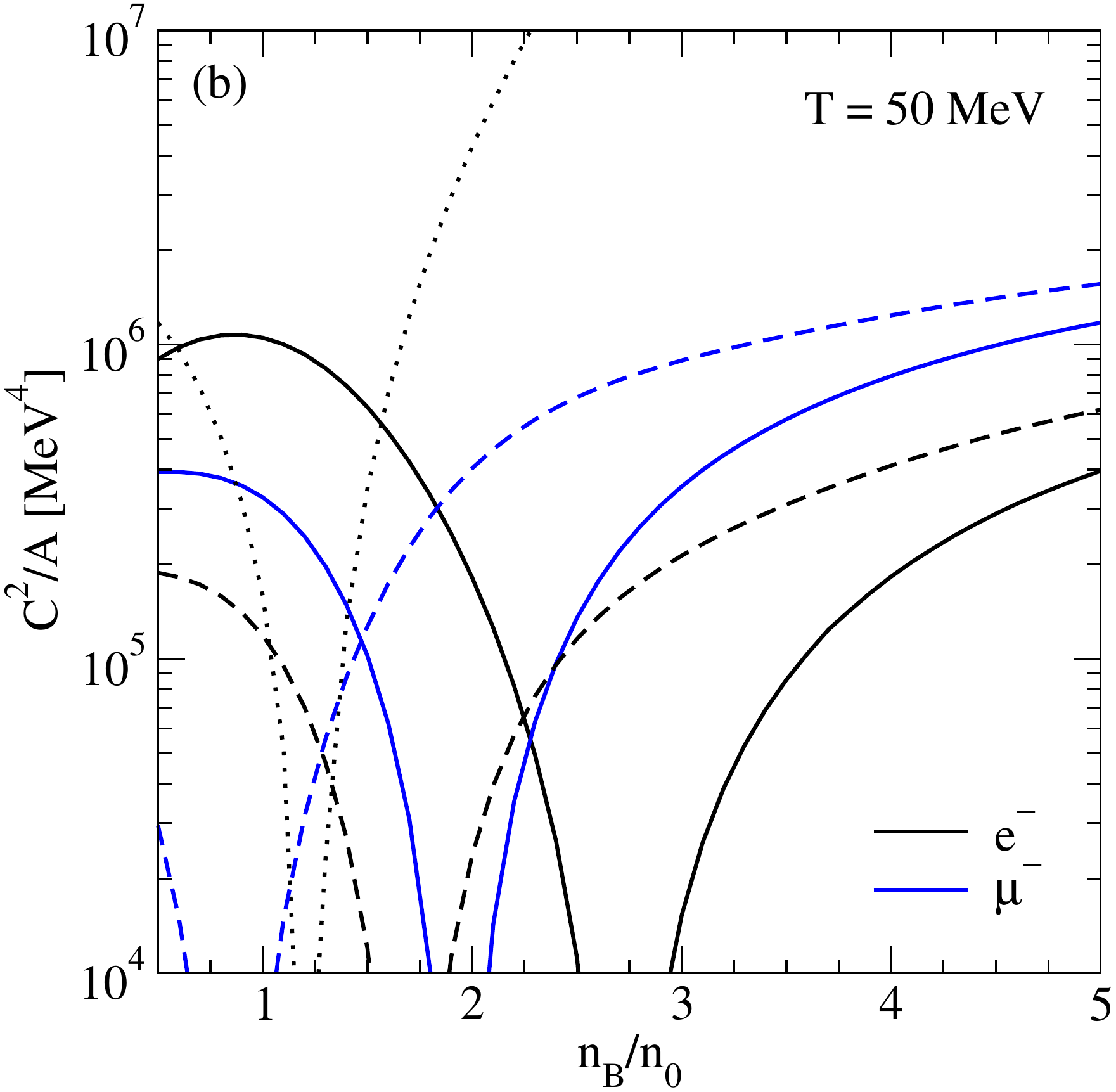}
\caption{
The susceptibilities $C^2_1/A_1$ and $C^2_2/A_2$ corresponding
to electronic and muonic bulk viscosities as functions of the baryon
density for neutrino-transparent matter (\textbf{a}) and neutrino-trapped matter
(\textbf{b}). The solid lines show isothermal susceptibilities, and the dashed
lines show the adiabatic susceptibilities. The dotted lines show the
nonrelativistic approximation to the isothermal
$C^2_1/A_1$~\cite{Alford2019b}.
}
\label{fig:C2A_dens}
\end{figure}

Figure~\ref{fig:C2A_dens} shows the susceptibilities $C_1^2/A_1$ and
$C_2^2/A_2$ (as computed in Ref.~\cite{Alford2021c}) which enter the
formulas of the partial bulk viscosities from electronic ($\zeta_e$)
and muonic ($\zeta_\mu$) Urca processes (note that $\zeta_\mu$ should
be obtained from Equation~\eqref{eq:zeta_slow1} by replacing $A_1\to A_2$,
$C_1\to C_2$ and $\gamma_e\to \gamma_\mu=\lambda_\mu A_2$).  Panels
(a) and (b) show the results for neutrino-transparent and
neutrino-trapped matter, respectively. The solid curves correspond to
isothermal, and the dashed lines---to adiabatic susceptibilities. In
the neutrino-transparent regime, the susceptibilities are sensitive to
the density and temperature only in the low-density region, where
the difference between isothermal and adiabatic susceptibilities is the
largest (\eg, the ratio of adiabatic and isothermal $C_1^2/A_1$ is
around 1.67 at $n_B=0.5n_0$ and $T=5$~MeV). In the high-temperature
regime of neutrino-trapped matter, there are special values
of the density where
$C_1^2/A_1$ and $C_2^2/A_2$, and therefore, also the partial bulk
viscosities $\zeta_e$ and $\zeta_\mu$ drop to zero as a result of the
subsystems of electrons and muons, respectively becoming scale-invariant
at those points. At those points the electron and muon fractions become
independent of the baryon density, as seen from the inset of Figure~\ref{fig:fractions}
(the small shift of the special point in the electron susceptibility from
the minimum of the electron fraction is a result
of the approximations made in the evaluation of the susceptibilities).

In contrast to the
low-temperature regime, the difference between isothermal and
adiabatic susceptibilities is significant in the neutrino-trapped
matter. Typically, the adiabaticity shifts the scale-invariant points to lower densities by about one nuclear density as compared to the
isothermal case. The muonic and electronic susceptibilities differ on
average by factors from 2 to 5 in both regimes (except the domain of
very low densities below the muon threshold in the
neutrino-transparent matter, and the vicinity of the scale-invariant point in
the case of neutrino-trapped matter). Next, comparing the panels (a)
and (b) of Figure~\ref{fig:C2A_dens}, we see that the susceptibilities
are roughly an order of magnitude larger in the neutrino-transparent
regime. We see also that the nonrelativistic approximation to nucleons
strongly overestimates the susceptibilities.

\begin{figure}[H]
\includegraphics[width=0.45\columnwidth,keepaspectratio]{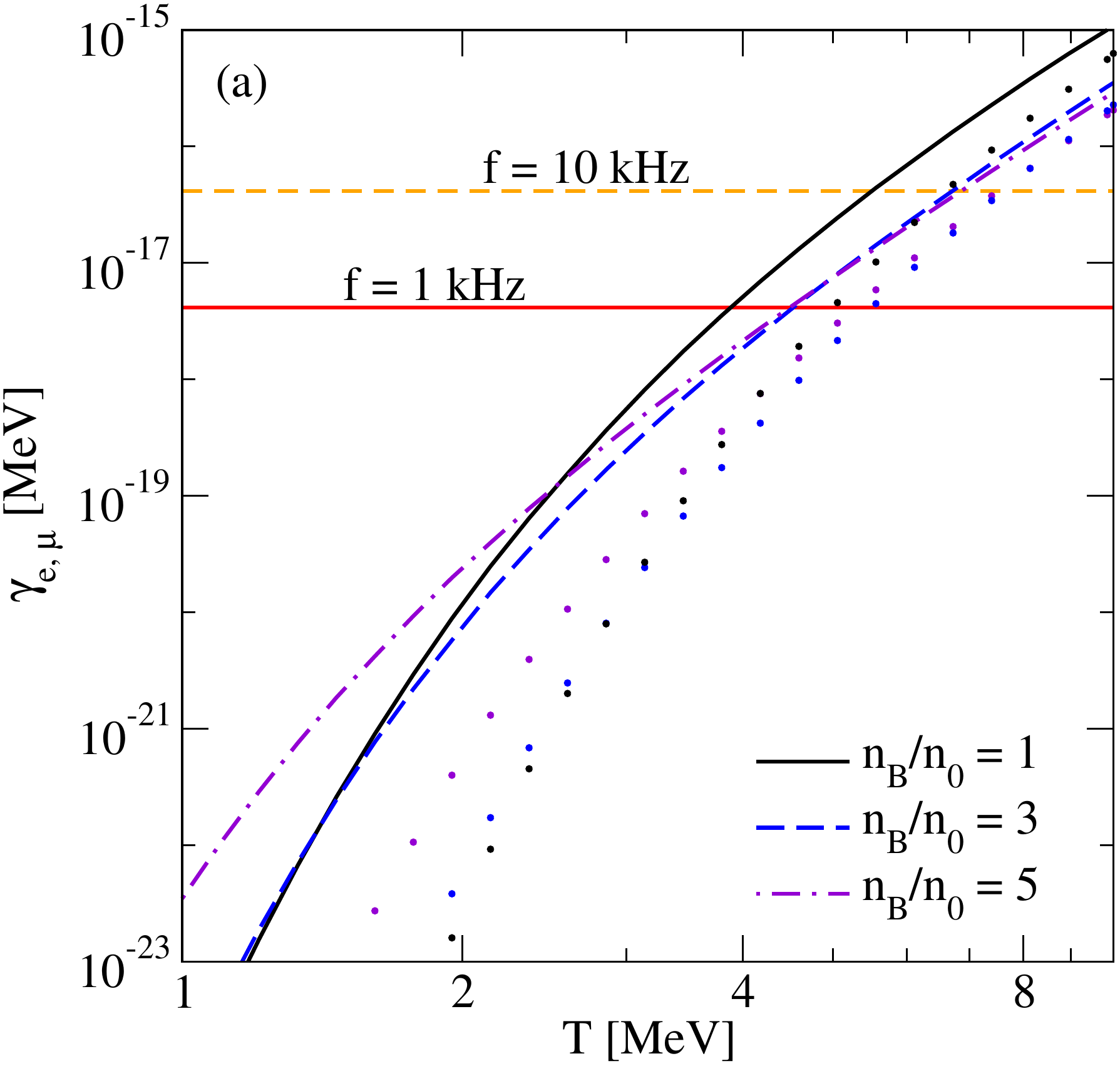}
\hspace{0.5cm}
\includegraphics[width=0.45\columnwidth,keepaspectratio]{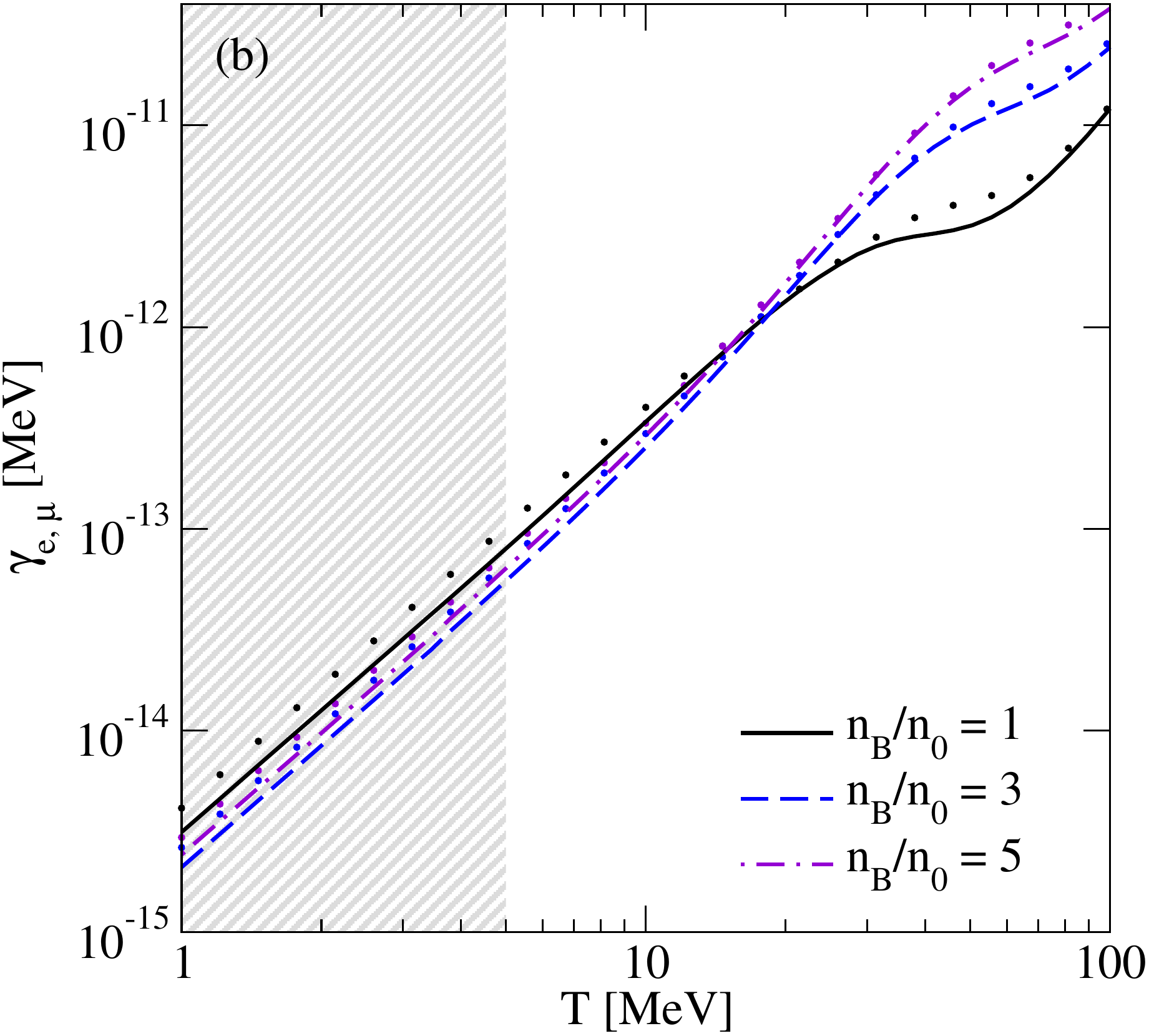}
\caption{The relaxation rates $\gamma_e$ (solid, dashed, and
dash-dotted lines) and $\gamma_\mu$ (dotted lines) of Urca processes
as functions of the temperature for fixed values of the density for
(\textbf{a}) neutrino-transparent matter; (\textbf{b}) neutrino-trapped matter, {where
the shaded area shows the extrapolation of the result to the
temperature regime $T\le 5$ MeV, where the trapping assumption fails.
}  The horizontal lines in panel (\textbf{a}) correspond to
the fixed values of oscillation frequency $f = 1$~kHz (solid line)
and $f = 10$~kHz (dashed
line).
}
\label{fig:gamma_e_temp}
\end{figure}

The relaxation rates $\gamma_e=\lambda_e A_1$ and $\gamma_\mu=\lambda_\mu A_2$
of electronic and muonic Urca processes, respectively, are shown in
Figure~\ref{fig:gamma_e_temp}. In the neutrino-transparent regime,
$\gamma_e$ and $\gamma_\mu$ cross the curves of the constant frequencies
$f\equiv\omega/2\pi=1$~kHz and $f\equiv\omega/2\pi=10$~kHz (1~kHz
corresponds to $4.14\cdot 10 ^{-18}$ MeV) at temperatures in the range
$3\leq T\leq 8$~MeV. Around these temperatures, the bulk viscosity of $npe\mu$
matter shows a resonant maximum. In the neutrino-trapped regime, the
relaxation rates are always higher than the typical oscillation frequencies,
and the bulk viscosity is independent of the oscillation frequency.

\subsection{Bulk Viscosity of {$npe \mu$} Matter in the Isothermal Case}
\label{sec:bulk_muons}



\begin{figure}[H]
\includegraphics[width=0.4\columnwidth, keepaspectratio]{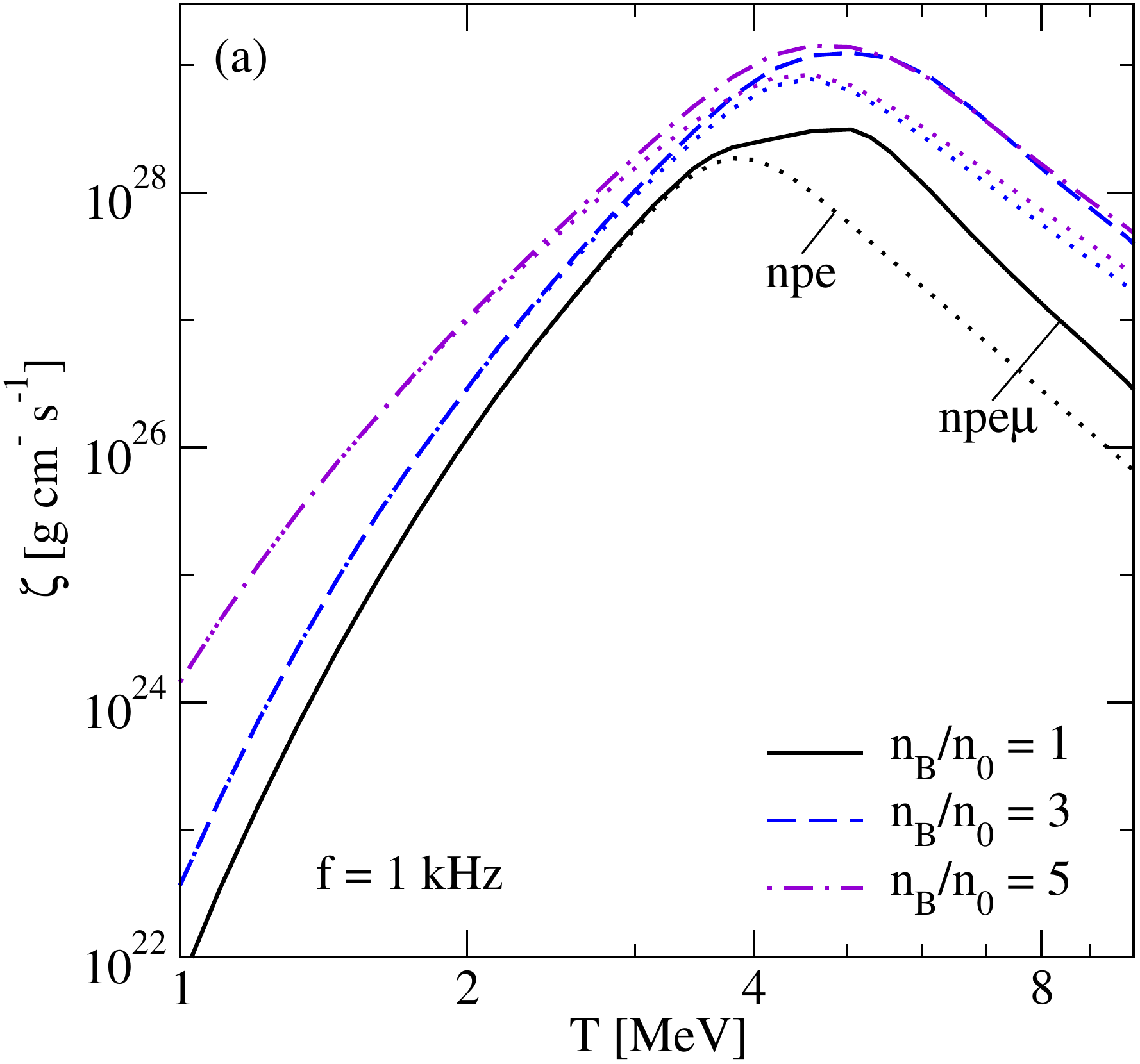}
\hspace{0.5cm}
\includegraphics[width=0.4\columnwidth, keepaspectratio]{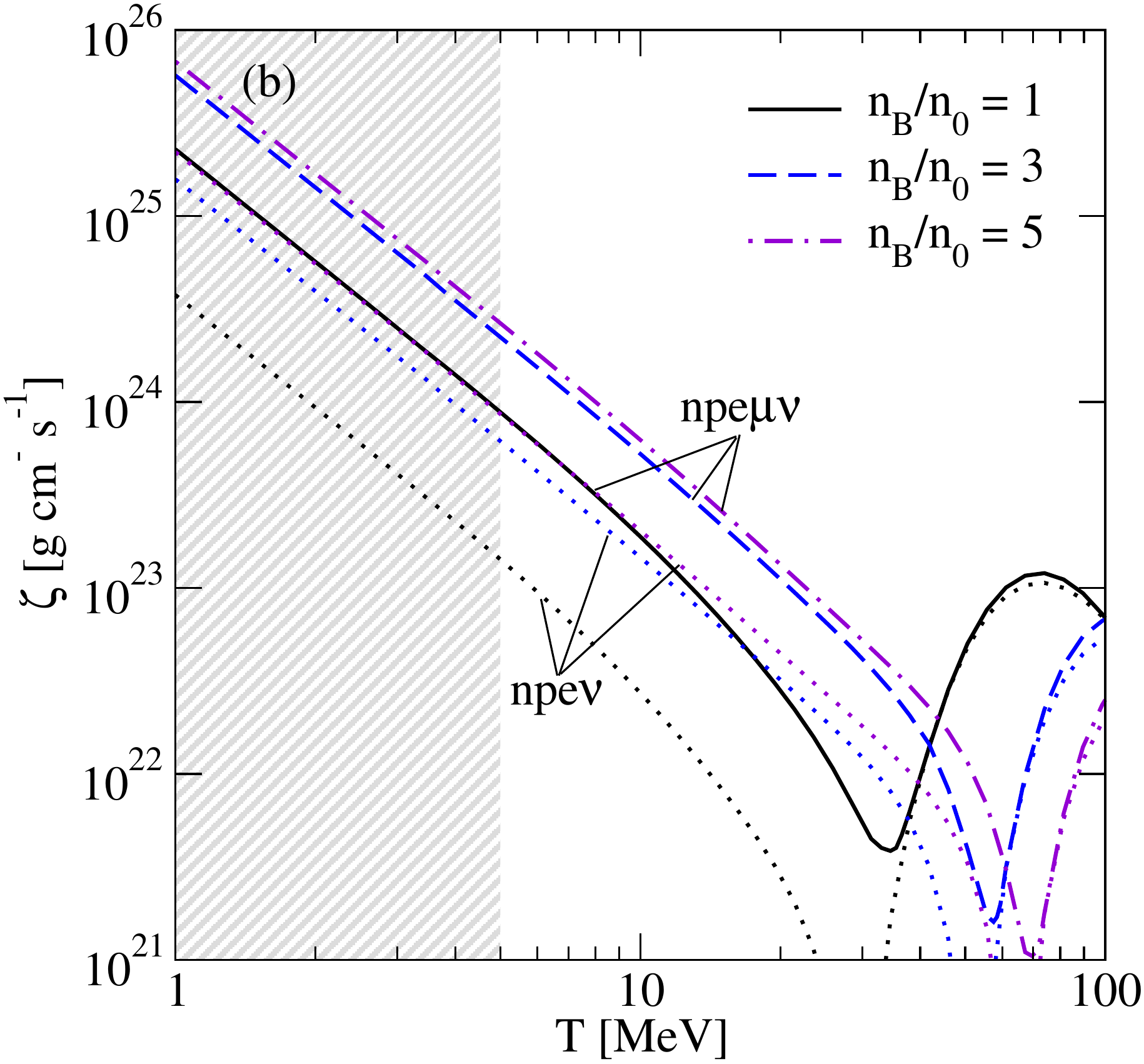}
\caption{The bulk viscosity of relativistic $npe\mu$ matter as a
function of the temperature for (\textbf{a}) neutrino-transparent matter at
$f=1$ kHz; (\textbf{b}) neutrino-trapped matter. The region $T\le 5$~MeV in
panel (\textbf{b}) is shaded because neutrinos are no longer trapped at those
temperatures. The dotted lines show the bulk viscosities of
relativistic $npe$ matter. }
\label{fig:zeta_slow_temp}
\end{figure}

The results for the bulk viscosity of relativistic $npe\mu$ matter
(computed with the isothermal susceptibilities) are shown in
Figure~\ref{fig:zeta_slow_temp}. Panel (a) shows
the results for neutrino-transparent matter at frequency $f=1$ kHz, which is typical
for density oscillations in neutron star mergers. At low temperatures,
where $\lambda_l A_j\ll\omega$, the bulk viscosity is given by the
sum of electronic and muonic partial viscosities, $\zeta=\zeta_e+
\zeta_\mu\propto \omega^{-2}$, see Equation~\eqref{eq:zeta_slow2}. In this
regime we have typically $\zeta_\mu\ll \zeta_e$, and the bulk viscosity
of $npe\mu$ matter practically coincides with that of $npe$ matter which
is shown in Figure~\ref{fig:zeta_slow_temp} by dotted lines. For the given
frequency the bulk viscosity of $npe\mu$ matter has a resonant maximum
at a temperature between the resonant temperatures $T_l$ of partial bulk
viscosities $\zeta_l$, where $\omega=\gamma_l(T_{l})$ with $l=\{e,\mu\}$.
The maximum of the bulk viscosity of $npe\mu$ matter is located at a
slightly higher temperature as compared to the bulk viscosity of $npe$
matter. At temperatures above the maximum, where $\lambda_l A_j\gg\omega$,
the bulk viscosity becomes frequency-independent. In this regime, the bulk
viscosity of $npe\mu$ matter exceeds the bulk viscosity of $npe$ matter
by factors between 2.5 and 8.

As in the neutrino-trapped matter, the equilibration rates are much
larger than the oscillation frequency, the bulk viscosity is given by the
frequency-independent Formula  \eqref{eq:zeta_slow3}. In the high-temperature
range we have mainly $(A_n+A_p) C_1\ll A_1 C_2$, $(A_n+A_p) C_2\ll A_2 C_1$,
$(A_n+A_p)^2\ll A_1A_2$, which allows to simplify Equation~\eqref{eq:zeta_slow3} to
\bea\label{eq:zeta_slow4}
\zeta \simeq \frac{\lambda_e (A_1 C_2)^2
+\lambda_\mu (A_2 C_1)^2 }
{\lambda_e \lambda_\mu (A_1A_2)^2}
=\frac{C_1^2 }{A_1}\frac{1}{\gamma_e}
+\frac{ C_2^2}{A_2}\frac{1}{\gamma_\mu}
=\zeta_e +\zeta_\mu.
\eea

As the susceptibilities, $C_1$ and $C_2$ cross zero at high temperatures, the
partial bulk viscosities drop to zero at { those points}
as well. The summed $\zeta$ will thus obtain a minimum at an intermediate temperature
but will remain finite at the minimum. The generic behavior of the bulk
viscosity of $npe\mu$ matter is similar to the one of $npe$ matter.
The former exceeds the latter by factors from 3 to 10 on the left side
of the minimum, whereas to the right side of the minimum the muonic
contribution to the bulk viscosity is negligible.

We also note that the nonrelativistic approximation would highly
overestimate the bulk viscosities in the whole temperature-density
regime considered because this approximation leads to an underestimate
of the equilibration rates and an overestimate of the susceptibilities.

\subsection{Damping of Density Oscillations}
\label{sec:damping}

\begin{figure}[H]
\includegraphics[width=0.45\columnwidth, keepaspectratio]{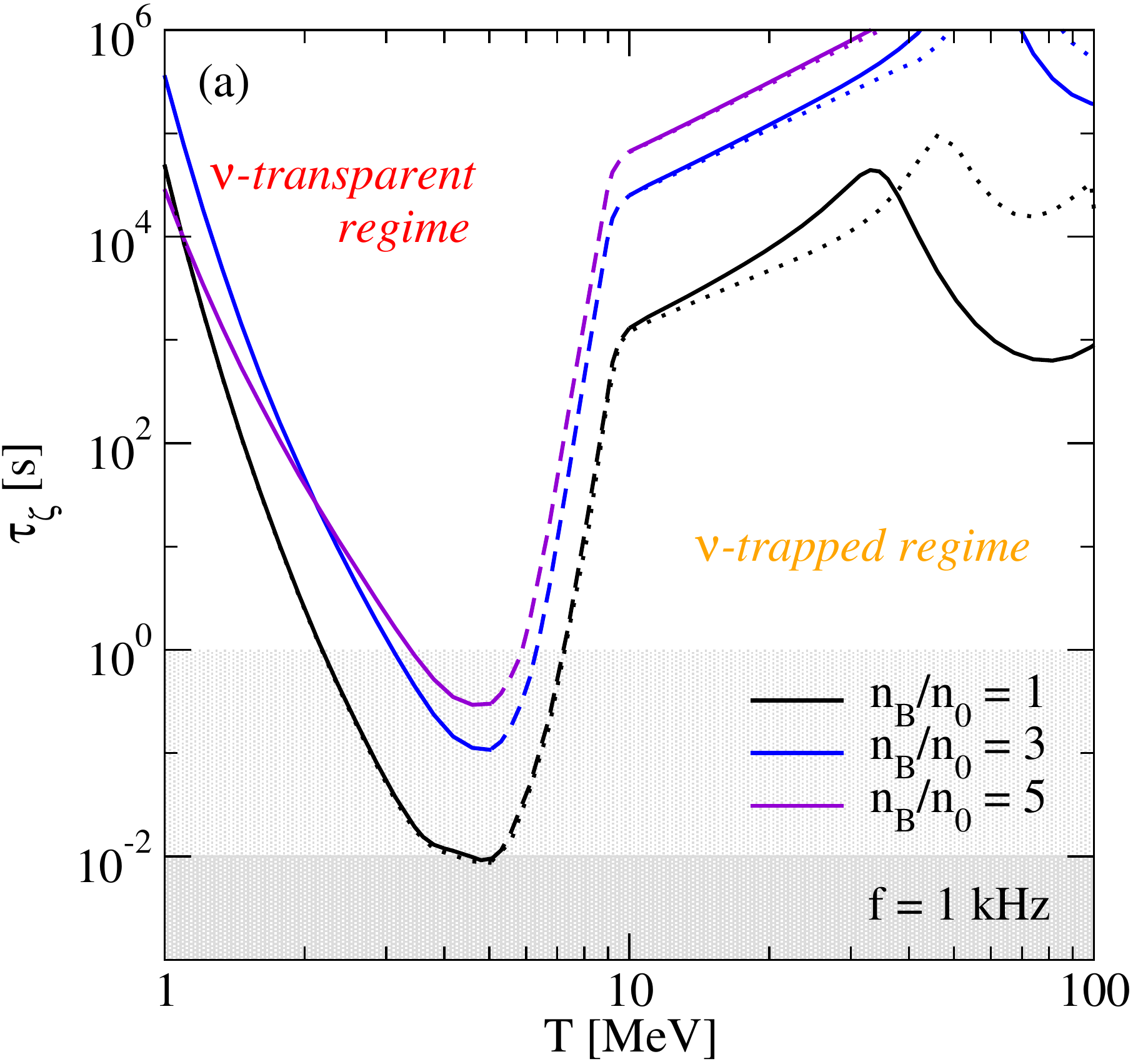}
\hspace{0.5cm}
\includegraphics[width=0.45\columnwidth, keepaspectratio]{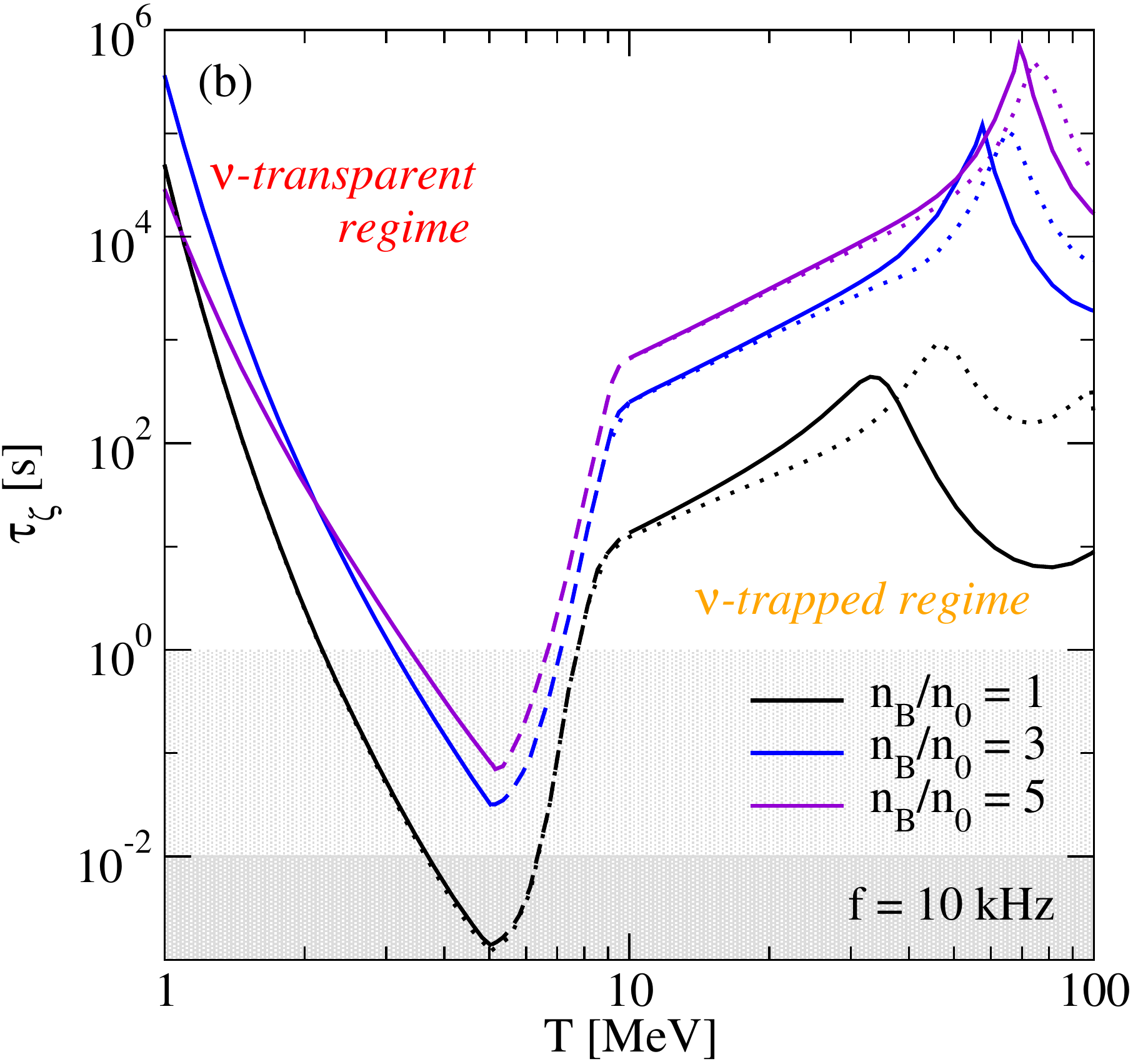}
\caption{The oscillation damping timescale as a function of temperature for various densities  for frequency fixed at (\textbf{a}) $f=1$ kHz; (\textbf{b}) $f=10$ kHz. The solid lines show the results
obtained with isothermal, and the dotted lines---with adiabatic susceptibilities.
The dashed lines interpolate between the results of neutrino-transparent and neutrino-trapped
regimes.}
\label{fig:tau_damp}
\end{figure}

Now we estimate the bulk viscous damping timescales of density
oscillations in relativistic $npe\mu$ matter. The damping
timescale is given by~\cite{Alford2018a,Alford2019a,Alford2020}
\bea\label{eq:damping_time}
\tau_{\zeta} =\frac{1}{9}\frac{Kn_B}{\omega^2\zeta},
\eea
where $\epsilon$ is the energy density of the system, and
\bea\label{eq:compress}
K=9n_B\frac{\partial^2\varepsilon}{\partial n_B^2}
\eea
is the incompressibility of nuclear matter. It depends weakly
on the temperature in both regimes of neutrino-transparent and
neutrino-trapped matter~\cite{Alford2019b}, therefore the temperature
dependence of the damping timescale is practically the inverse temperature
dependence of the bulk viscosity. Figure~\ref{fig:tau_damp} shows the
damping timescale for two oscillation frequencies: panel (a) with $f=1$~kHz,
and panel (b) with $f=10$~kHz. In each of the panels, we combined the
results of neutrino-transparent ($1\leq T\leq 5$~MeV) and neutrino-trapped
matter ($10\leq T\leq 100$~MeV). For intermediate temperatures $5\leq T\leq
10$~MeV the results are extrapolated between these two regimes with dashed lines.
The damping timescale attains its minimum around $T=5$~MeV,
with its value being inversely proportional to the frequency. In the low-temperature
regime (to the left side of the minimum) the damping timescale is
frequency-independent but becomes inversely proportional to the
square of $\omega$ in the neutrino-trapped regime as the bulk
viscosity is independent of the oscillation frequency there.



The shaded areas in Figure~\ref{fig:tau_damp} separate the
temperature-density range where the damping timescale becomes smaller
than the early ($\simeq$10\,ms, dark shaded areas) and long-term
($\simeq$1 s, lightly shaded areas) evolution timescales of post-merger
object, respectively. For a typical frequency $f=1$~kHz, the bulk
viscous damping is efficient in short-living remnants only at low
densities $n_B\leq n_0$ in the temperature range $4\leq T\leq
6$~MeV. At higher densities, the damping could be relevant during the
long-term evolution only. For higher frequencies, there is a larger
domain of densities and temperatures where the damping timescales
reach down to the short-term evolution timescale of BNS mergers. For
example, for $f=10$~kHz the short-term damping is efficient at
densities $n_B\leq 2n_0$ and for temperatures between
$3\leq T\leq 7$~MeV.  The dynamics of long-living remnants would be
affected by the bulk viscosity for a wider temperature-density range,
typically $2\leq T\leq 10$~MeV and $n_B\leq 5n_0$.

For the sake of completeness, we show also the damping timescales computed with
adiabatic susceptibilities with dotted lines in Figure~\ref{fig:tau_damp}.
Note that the bulk viscosities and the damping timescales
calculated in Ref.~\cite{Alford2019a} used the adiabatic susceptibilities
and compressibilities. It was found that the bulk viscosities and the
damping timescales for adiabatic and isothermal oscillations in the
neutrino-transparent matter differ by a factor of around 2 in the regime
where the bulk viscous damping is efficient in the post-merger dynamics.
This is fully consistent with our findings, see
Figure~\ref{fig:tau_damp}. As expected, the adiabaticity modifies the results significantly only in the
high-temperature regime, where it can increase $\tau_\zeta$ by more than an
order of magnitude. However, in the high-temperature regime of neutrino-trapped
matter the Urca-process-driven bulk viscosity is not sufficiently large to
affect the evolution of BNS mergers.

\section{Conclusions}
\label{sec:conclusions}

In this work, we provided a brief review of our work on the
Urca-process-driven bulk viscosity of relativistic $npe\mu$ matter in
the parameter range relevant to binary neutron star mergers. We
focused on the semi-leptonic Urca processes as well as leptonic
processes in $npe\mu$ matter in two different regimes of interest:
(a) neutrino-transparent regime where $T\leq T_{\rm tr}$ with
$T_{\rm tr}\simeq 5\div 10$~MeV being the neutrino-trapping
temperature; and (b) neutrino-trapped regime at $T\geq T_{\rm tr}$.
Along with the results for the bulk viscosity in the
neutrino-trapped regime obtained earlier in Ref.~\cite{Alford2021c},
we showed novel results for the neutrino-transparent matter as well
as some results pertaining to the case of isentropic instead of
isothermal oscillations.

Our main observations can be summarized as follows:
\begin{enumerate}[labelsep=5mm]
\item[(a)] We observe that the leptonic reactions proceed much slower than the
Urca process in the entire temperature-density range.
In the neutrino-transparent matter the dominant leptonic reaction
is the muon decay, whereas
in the neutrino-trapped regime, the dominant leptonic reactions
are the neutrino and antineutrino absorption processes.

\item[(b)] As a result, the bulk viscosity of $npe\mu$ matter can be computed assuming
that the leptonic processes are frozen. Qualitatively, the bulk viscosity of
$npe\mu$ matter exceeds that of $npe$ matter by factors from 2.5 to 8 above
the maximum temperature in the $\nu$-transparent matter, and by factors from
1 to 10 in the $\nu$-trapped matter.

\item[(c)] The bulk viscosity features its resonant maximum at a temperature
where the average $\beta$-relaxation rate of electronic and muonic Urca
processes coincides with the angular frequency of density oscillations.
This resonant maximum appears around $T\simeq 5$~MeV where the matter is
still transparent to neutrinos. At higher temperatures, bulk viscosity drops
rapidly once the matter enters the neutrino-trapped regime. There
appear sharp minima in the bulk viscosity at $T\geq 30$\,MeV where the
lepton fractions become independent of the density.

\item[(d)] \textls[-5]{Using our results for the bulk viscosity we estimate the bulk viscous
damping timescales of density oscillations in neutron star mergers. We
find that for typical oscillation frequencies $1\leq f\leq 10$~kHz there
is a finite temperature-density range where the bulk viscous dissipation
can affect the short-term evolution ($\simeq$10~ms) of BNS mergers
significantly. The damping timescale features a minimum at $T\simeq 5$~MeV
with minimum values of the order of ms at very low densities $n_B\leq n_0$.
At higher densities, the damping timescales of density oscillations are larger
and can affect the post-merger evolution only on a long-time scale. At high
temperatures where neutrinos are trapped in the matter, the damping timescales
are much longer; therefore, the Urca processes are not the dominant
channels by which
to damp the density oscillations in BNS mergers.}
\end{enumerate}

\vspace{6pt}

\authorcontributions{The authors contributed equally to this research. {All authors} 
have read and agreed to the published version of the manuscript. }

\funding{The research of M.A. was funded by the U.S. Department of
Energy, Office of Science, Office of Nuclear Physics under Award
Number No.~DE-FG02-05ER41375.  The research of A.H. and A.S. was
funded by the Volkswagen Foundation (Hannover, Germany) grant No.~96
839 and the European COST Action ``PHAROS'' (CA16214).  The research
of A.S. was funded by Deutsche Forschungsgemeinschaft Grant No. SE
1836/5-2 and the Polish NCN Grant No. 2020/37/B/ST9/01937 at Wroclaw
University.}


\dataavailability{{Not applicable.} 
}

\conflictsofinterest{The authors declare no conflict of interest.}

\begin{adjustwidth}{-\extralength}{0cm}

\reftitle{References}



\begin{thebibliography}{999}

\bibitem[{Endrizzi} \em{et~al.}(2018){Endrizzi}, {Logoteta}, {Giacomazzo},
{Bombaci}, {Kastaun}, and {Ciolfi}]{Endrizzi2018}
{Endrizzi}, A.; {Logoteta}, D.; {Giacomazzo}, B.; {Bombaci}, I.; {Kastaun}, W.;
{Ciolfi}, R.
\newblock {Effects of chiral effective field theory equation of state on binary
neutron star mergers}.
\newblock {\em \prd} {\bf 2018}, {\em 98},~043015.
\newblock  [\href{http://doi.org/10.1103/PhysRevD.98.043015}{CrossRef}]

\bibitem[{Most} \em{et~al.}(2019){Most}, {Papenfort}, and {Rezzolla}]{Most2019}
{Most}, E.R.; {Papenfort}, L.J.; {Rezzolla}, L.
\newblock {Beyond second-order convergence in simulations of magnetized binary
neutron stars with realistic microphysics}.
\newblock {\em \mnras} {\bf 2019}, {\em 490},~3588--3600.
\newblock  [\href{http://dx.doi.org/10.1093/mnras/stz2809}{CrossRef}]

\bibitem[{Ciolfi} \em{et~al.}(2019){Ciolfi}, {Kastaun}, {Kalinani}, and
{Giacomazzo}]{Ciolfi2019}
{Ciolfi}, R.; {Kastaun}, W.; {Kalinani}, J.V.; {Giacomazzo}, B.
\newblock {First 100 ms of a long-lived magnetized neutron star formed in a
binary neutron star merger}.
\newblock {\em \prd} {\bf 2019}, {\em 100},~023005.
\newblock  [\href{http://dx.doi.org/10.1103/PhysRevD.100.023005}{CrossRef}]

\bibitem[{Tsokaros} \em{et~al.}(2019){Tsokaros}, {Ruiz}, {Paschalidis},
{Shapiro}, and {Ury{\={u}}}]{Tsokaros2019}
{Tsokaros}, A.; {Ruiz}, M.; {Paschalidis}, V.; {Shapiro}, S.L.; {Ury{\={u}}},
K.
\newblock {Effect of spin on the inspiral of binary neutron stars}.
\newblock {\em \prd} {\bf 2019}, {\em 100},~024061.
\newblock  [\href{http://dx.doi.org/10.1103/PhysRevD.100.024061}{CrossRef}]

\bibitem[Most \em{et~al.}(2022)Most, Haber, Harris, Zhang, Alford, and
Noronha]{Most:2022yhe}
Most, E.R.; Haber, A.; Harris, S.P.; Zhang, Z.; Alford, M.G.; Noronha, J.
\newblock {Emergence of microphysical viscosity in binary neutron star
post-merger dynamics.} \emph{arXiv }{\bf 2022}, arXiv:2207.00442.

\bibitem[{Sawyer} and {Soni}(1979)]{Sawyer1979ApJ}
{Sawyer}, R.F.; {Soni}, A.
\newblock {Transport of neutrinos in hot neutron-star matter}.
\newblock {\em \apj} {\bf 1979}, {\em 230},~859--869.
\newblock  [\href{http://dx.doi.org/10.1086/157146}{CrossRef}]

\bibitem[{Sawyer}(1980)]{Sawyer1980ApJ}
{Sawyer}, R.F.
\newblock {Damping of neutron star pulsations by weak interaction processes}.
\newblock {\em \apj} {\bf 1980}, {\em 237},~187--197.
\newblock  [\href{http://dx.doi.org/10.1086/157858}{CrossRef}]

\bibitem[{Sawyer}(1989)]{Sawyer1989}
{Sawyer}, R.F.
\newblock {Bulk viscosity of hot neutron-star matter and the maximum rotation \
rates of neutron stars}.
\newblock {\em \prd} {\bf 1989}, {\em 39},~3804--3806.
\newblock  [\href{http://dx.doi.org/10.1103/PhysRevD.39.3804}{CrossRef}]

\bibitem[{Haensel} and {Schaeffer}(1992)]{Haensel1992PhRvD}
{Haensel}, P.; {Schaeffer}, R.
\newblock {Bulk viscosity of hot-neutron-star matter from direct URCA
processes}.
\newblock {\em \prd} {\bf 1992}, {\em 45},~4708--4712.
\newblock  [\href{http://dx.doi.org/10.1103/PhysRevD.45.4708}{CrossRef}]

\bibitem[{Haensel} \em{et~al.}(2000){Haensel}, {Levenfish}, and
{Yakovlev}]{Haensel2000}
{Haensel}, P.; {Levenfish}, K.P.; {Yakovlev}, D.G.
\newblock {Bulk viscosity in superfluid neutron star cores. I. Direct Urca
processes in npemu matter}.
\newblock {\em \aap} {\bf 2000}, {\em 357},~1157--1169.

\bibitem[{Haensel} \em{et~al.}(2001){Haensel}, {Levenfish}, and
{Yakovlev}]{Haensel2001}
{Haensel}, P.; {Levenfish}, K.P.; {Yakovlev}, D.G.
\newblock {Bulk viscosity in superfluid neutron star cores. II. Modified Urca
processes in npe mu matter}.
\newblock {\em \aap} {\bf 2001}, {\em 372},~130--137.
\newblock  [\href{http://dx.doi.org/10.1051/0004-6361:20010383}{CrossRef}]

\bibitem[Haensel \em{et~al.}(2002)Haensel, Levenfish, and
Yakovlev]{Haensel2002}
Haensel, P.; Levenfish, K.; Yakovlev, D.
\newblock {Bulk viscosity in superfluid neutron star cores. III. Effects of
sigma-hyperons}.
\newblock {\em \aap} {\bf 2002}, {\em 381},~1080--1089.
\newblock  [\href{http://dx.doi.org/10.1051/0004-6361:20011532}{CrossRef}]

\bibitem[{Dong} \em{et~al.}(2007){Dong}, {Su}, and {Wang}]{Dong2007}
{Dong}, H.; {Su}, N.; {Wang}, Q.
\newblock {Bulk viscosity in nuclear and quark matter}.
\newblock {\em J. Phys. Nucl. Phys.} {\bf 2007}, {\em
34},~S643--S646.
\newblock  [\href{http://dx.doi.org/10.1088/0954-3899/34/8/S63}{CrossRef}]

\bibitem[{Alford} \em{et~al.}(2010){Alford}, {Mahmoodifar}, and
{Schwenzer}]{Alford2010JPhG}
{Alford}, M.G.; {Mahmoodifar}, S.; {Schwenzer}, K.
\newblock {Large amplitude behavior of the bulk viscosity of dense matter}.
\newblock {\em J. Phys. Nucl. Phys.} {\bf 2010}, {\em
37},~125202.
\newblock  [\href{http://dx.doi.org/10.1088/0954-3899/37/12/125202}{CrossRef}]

\bibitem[Alford and Good(2010)]{Alford:2010jf}
Alford, M.G.; Good, G.
\newblock {Leptonic contribution to the bulk viscosity of nuclear matter}.
\newblock {\em Phys. Rev.} {\bf 2010}, {\em C82},~055805.
\newblock  [\href{http://dx.doi.org/10.1103/PhysRevC.82.055805}{CrossRef}]

\bibitem[{Kolomeitsev} and {Voskresensky}(2015)]{Kolomeitsev2015}
{Kolomeitsev}, E.E.; {Voskresensky}, D.N.
\newblock {Viscosity of neutron star matter and r-modes in rotating pulsars}.
\newblock {\em \prc} {\bf 2015}, {\em 91},~025805.
\newblock  [\href{http://dx.doi.org/10.1103/PhysRevC.91.025805}{CrossRef}]

\bibitem[{Alford} and {Harris}(2019)]{Alford2019a}
{Alford}, M.G.; {Harris}, S.P.
\newblock {Damping of density oscillations in neutrino-transparent nuclear
matter}.
\newblock {\em \prc} {\bf 2019}, {\em 100},~035803.
\newblock  [\href{http://dx.doi.org/10.1103/PhysRevC.100.035803}{CrossRef}]

\bibitem[{Alford} \em{et~al.}(2019){Alford}, {Harutyunyan}, and
{Sedrakian}]{Alford2019b}
{Alford}, M.; {Harutyunyan}, A.; {Sedrakian}, A.
\newblock {Bulk viscosity of baryonic matter with trapped neutrinos}.
\newblock {\em \prd} {\bf 2019}, {\em 100},~103021.
\newblock  [\href{http://dx.doi.org/10.1103/PhysRevD.100.103021}{CrossRef}]

\bibitem[{Alford} and {Haber}(2021)]{Alford2021a}
{Alford}, M.G.; {Haber}, A.
\newblock {Strangeness-changing rates and hyperonic bulk viscosity in neutron
star mergers}.
\newblock {\em \prc} {\bf 2021}, {\em 103},~045810.
\newblock  [\href{http://dx.doi.org/10.1103/PhysRevC.103.045810}{CrossRef}]

\bibitem[{Alford} \em{et~al.}(2021){Alford}, {Harutyunyan}, and
{Sedrakian}]{Alford2021c}
{Alford}, M.; {Harutyunyan}, A.; {Sedrakian}, A.
\newblock {Bulk viscosity from Urca processes: {\ensuremath{npe\mu}} matter in
the neutrino-trapped regime}.
\newblock {\em \prd} {\bf 2021}, {\em 104},~103027.
\newblock  [\href{http://dx.doi.org/10.1103/PhysRevD.104.103027}{CrossRef}]

\bibitem[{Lalazissis} \em{et~al.}(2005){Lalazissis}, {Nik{\v s}i{\'c}}, and
{Vretenar}]{Lalazissis2005}
Lalazissis, G.A.; Nikšić, T.; Vretenar, D.; Ring, P.
\newblock {New relativistic mean-field interaction with density-dependent meson-nucleon couplings}.
\newblock {\em \prc} {\bf 2005}, {\em 71},~024312.
\newblock  [\href{http://dx.doi.org/10.1103/PhysRevC.71.024312}{CrossRef}]

\bibitem[{Reed} \em{et~al.}(2021){Reed}, {Fattoyev}, {Horowitz} and
{Piekarewicz}]{Reed_2021}
Reed, B.T.; Fattoyev, F.J.; Horowitz, C.J.; Piekarewicz, J.
\newblock {Implications of PREX-2 on the Equation of State of Neutron-Rich Matter}.
\newblock {\em \prl} {\bf 2021}, {\em 126},~172503. [\href{http://dx.doi.org/10.1103/PhysRevLett.126.172503}{CrossRef}] [\href{http://www.ncbi.nlm.nih.gov/pubmed/33988426}{PubMed}]


\bibitem[{Reinhard} \em{et~al.}(2021)]{Reinhard_2021}
{Reinhard}, P.-G.; {Roca-Maza}, X.; {Nazarewicz}, W.
\newblock {Information Content of the Parity-Violating Asymmetry in $^{208}$Pb}.
\newblock {\em \prl} {\bf 2021}, {\em 127},~232501.


\bibitem[{Alford} and {Harris}(2018)]{Alford2018b}
{Alford}, M.G.; {Harris}, S.P.
\newblock {{$\beta$} equilibrium in neutron-star mergers}.
\newblock {\em \prc} {\bf 2018}, {\em 98},~065806.
\newblock  [\href{http://dx.doi.org/10.1103/PhysRevC.98.065806}{CrossRef}]

\bibitem[{Baiotti}(2019)]{Baiotti2019}
{Baiotti}, L.
\newblock {Gravitational waves from neutron star mergers and their relation to
the nuclear equation of state}.
\newblock {\em Prog. Part. Nucl. Phys.} {\bf 2019}, {\em
109},~103714.
\newblock  [\href{http://dx.doi.org/10.1016/j.ppnp.2019.103714}{CrossRef}]

\bibitem[Greiner and M{\"u}ller(2000)]{Greiner2000gauge}
Greiner, W.; M{\"u}ller, B.
\newblock {\em Gauge Theory of Weak Interactions}; Physics and Astronomy Online
Library; Springer: {Berlin/Heidelberg, Germany,} 
2000.

\bibitem[Guo \em{et~al.}(2020)Guo, Mart\'\i{}nez-Pinedo, Lohs, and
Fischer]{Guo:2020tgx}
Guo, G.; Mart\'\i{}nez-Pinedo, G.; Lohs, A.; Fischer, T.
\newblock {Charged-Current Muonic Reactions in Core-Collapse Supernovae}.
\newblock {\em Phys. Rev. D} {\bf 2020}, {\em 102},~023037.
\newblock  [\href{http://dx.doi.org/10.1103/PhysRevD.102.023037}{CrossRef}]

\bibitem[{Colucci} and {Sedrakian}(2013)]{Colucci2013}
{Colucci}, G.; {Sedrakian}, A.
\newblock {Equation of state of hypernuclear matter: Impact of hyperon-scalar-\
meson couplings}.
\newblock {\em \prc} {\bf 2013}, {\em 87},~055806.
\newblock  [\href{http://dx.doi.org/10.1103/PhysRevC.87.055806}{CrossRef}]

\bibitem[{Alford} \em{et~al.}(2021){Alford}, {Haber}, {Harris}, and
{Zhang}]{Alford2021b}
{Alford}, M.G.; {Haber}, A.; {Harris}, S.P.; {Zhang}, Z.
\newblock {Beta Equilibrium Under Neutron Star Merger Conditions}.
\newblock {\em Universe} {\bf 2021}, {\em 7},~399.
\newblock  [\href{http://dx.doi.org/10.3390/universe7110399}{CrossRef}]

\bibitem[{Alford} \em{et~al.}(){Alford}, {Harutyunyan}, and
{Sedrakian}]{Alford2022preprint}
{Alford}, M.; {Harutyunyan}, A.; {Sedrakian}, A.
\newblock {Bulk viscosity from Urca processes:  {\ensuremath{npe\mu}} matter in
the neutrino-transparent regime}.   {\ 2022}, \emph{preprint}.



\bibitem[{Alford} \em{et~al.}(2018){Alford}, {Bovard}, {Hanauske}, {Rezzolla},
and {Schwenzer}]{Alford2018a}
{Alford}, M.G.; {Bovard}, L.; {Hanauske}, M.; {Rezzolla}, L.; {Schwenzer}, K.
\newblock {Viscous Dissipation and Heat Conduction in Binary Neutron-Star
Mergers}.
\newblock {\em \prl} {\bf 2018}, {\em 120},~041101.
\newblock  [\href{http://dx.doi.org/10.1103/PhysRevLett.120.041101}{CrossRef}]

\bibitem[Alford \em{et~al.}(2020)Alford, Harutyunyan, and
Sedrakian]{Alford2020}
Alford, M.; Harutyunyan, A.; Sedrakian, A.
\newblock {Bulk Viscous Damping of Density Oscillations in Neutron Star
Mergers}.
\newblock {\em Particles} {\bf 2020}, {\em 3},~500--517.
\newblock  [\href{http://dx.doi.org/10.3390/particles3020034}{CrossRef}]

\end{thebibliography}
%


%
%
%
\end{adjustwidth}
\end{document}